\documentclass[twocolumn,showpacs,preprintnumbers,amsmath,amssymb,superscriptaddress]{revtex4}
\usepackage{graphicx}% Include figure files
\usepackage{dcolumn}% Align table columns on decimal point
\usepackage{bm}% bold math
\usepackage[chatter]{rotating}

\def\bea{\begin{eqnarray}}
\def\eea{\end{eqnarray}}

\begin{document}

%\preprint{Version 2.2}

\title{Challenging the utility of third-order azimuth harmonics in the description of ultra-relativistic heavy-ion collisions}

\affiliation{Department of Physics, The University of Texas at Austin, Austin, Texas 78712 USA}
\affiliation{CENPA 354290, University of Washington, Seattle, Washington 98195 USA}
\author{R. L. Ray}\affiliation{Department of Physics, The University of Texas at Austin, Austin, Texas 78712 USA}
\author{D. J. Prindle}\affiliation{CENPA 354290, University of Washington, Seattle, Washington 98195 USA}
\author{T. A. Trainor}\affiliation{CENPA 354290, University of Washington, Seattle, Washington 98195 USA}

%%%%%%%%%%%%%%%%%%%%%%%%%%%%%%%%%%%%%%%
\date{\today}

\begin{abstract}
In recent years it has become conventional practice to include higher-order cylindrical harmonics in the phenomenological description of two-particle angular correlations from ultra-relativistic heavy-ion collisions. These model elements, whose dependence on relative azimuth angle has the form $\cos[m(\phi_1-\phi_2)]$ where $m > 2$, were introduced to support a hydrodynamic flow interpretation of the same-side ($|\phi_1-\phi_2| < \pi/2$) 2D peak in the correlations. Previous studies have shown that the $m > 2$ harmonics are not required by the data, that they destabilize the fitting models, and that their net effect is to decompose the same-side peak into two components, one being dependent on and the other being independent of relative pseudorapidity. Thus we are lead to question whether descriptions of angular correlation data including higher-order harmonics inform our understanding of the same-side peak or heavy-ion collisions in general. Results from analysis of two-dimensional angular correlation data from the Relativistic Heavy-Ion Collider (RHIC) and the Large Hadron Collider (LHC) show that the RHIC data do not exclude a single-Gaussian hypothesis for the same-side peak. We find that the net effect of including the $m = 3$ harmonic or azimuth sextupole in the fitting model is the inclusion of small non-Gaussian dependence in the mathematical description of the same-side peak. Those non-Gaussian effects are systematically insignificant and can be accommodated by minor perturbations to the same-side 2D Gaussian peak model, which act locally at small relative azimuth. We also demonstrate that the 0-1\% 2D angular correlation data for 2.76 TeV Pb+Pb collisions from ATLAS, which display an away-side double peak on azimuth, do not require a sextupole and exclude a positive same-side sextupole.
\end{abstract}


\maketitle

\section{Introduction}
\label{SecI}

One of the more interesting observations to emerge from the study of two-particle angular correlation data from heavy-ion collisions at the RHIC and the LHC is the appearance of a two-dimensional (2D) peak at small relative azimuth ($\phi$) which significantly increases in amplitude and in width along relative pseudorapidity ($\eta$) for more-central collisions \cite{aya,axialCI,Joern,atlas,cms,alice}. For minimum-bias proton + proton (p+p) collisions and for Au + Au collisions from peripheral to mid-centrality (50\% of fractional cross section) at RHIC this correlation peak structure is consistent with perturbative quantum chromodynamics (QCD) predictions for minimum-bias jets (no lower momentum cut) assuming binary nucleon-nucleon collision scaling \cite{axialCI,Tomjetfrag,Tommodfrag}. Beginning at mid-centrality for Au+Au collisions at $\sqrt{s_{\rm NN}}$ = 62 and 200~GeV the amplitude and the width on relative pseudorapidity abruptly increase more rapidly with centrality (a {\em sharp transition} in the parameter trends defined in \cite{axialCI}) leading to a broad, 2D peaked structure that extends beyond the tracking acceptance. Some authors ({\em e.g.} Refs.~\cite{Joern,PHOBOSridge}) interpret this structure as the combination of an $\eta$-independent ``ridge'' on azimuth plus a reduced 2D peak. A similar same-side (relative azimuth $< \pi/2$) 2D peaked structure was reported by the PHOBOS collaboration~\cite{PHOBOSridge} and for the 2.76 TeV Pb+Pb collision data and angular correlations by the ALICE~\cite{alice}, CMS~\cite{cms} and ATLAS~\cite{atlas} collaborations at the LHC.

In 2010 Alver and Roland \cite{AlverRoland} used a truncated Fourier cosine series ($\sum_m A_m \cos[m(\phi_1 - \phi_2)]$, $m \leq 3$) to fit the azimuth projection of 2D angular correlation data from PHOBOS~\cite{PHOBOSridge,PHOBOSdata} and STAR~\cite{STARdata} for large relative pseudorapidity. Using a Monte Carlo Glauber model for Au+Au collisions they obtained non-zero initial-state eccentricities $\varepsilon_2$ and $\varepsilon_3$ for the participant nucleon positions on the transverse plane perpendicular to the beam direction. Then, using Monte Carlo events from AMPT~\cite{AMPT} they found that the final-state $v_2$ and $v_3$ were linearly proportional to the respective nucleon participant $\varepsilon_2$ and $\varepsilon_3$ when the $v_m$ were calculated relative to the $m^{\rm th}$-order participant plane. They claimed that this linear relationship for $v_3$ is evidence of {\em triangular} flow in the AMPT transport mechanism. They went on to suggest that the pseudorapidity-elongated same-side peak in the 2D angular correlations from PHOBOS~\cite{PHOBOSridge,PHOBOSdata} and STAR~\cite{STARdata} is partially caused by, and perhaps dominated by triangular flow.  

Use of higher-order azimuth harmonics ($m > 2$) in phenomenological descriptions of angular correlation data is now ubiquitous in the literature. Recent 2D correlation data from ATLAS~\cite{atlas} and 1D azimuth correlation data from ALICE~\cite{alice} for 2.76 TeV Pb+Pb most-central collisions (0-1\% and 0-5\% of the total cross section) reveal an away-side double-peaked structure on azimuth which is interpreted to suggest a third-order harmonic contribution.

In this paper we challenge the introduction of ``higher harmonics'' to the 2D angular correlation data description. We focus on the $m = 3$ cylindrical harmonic (sextupole); the $m > 3$ terms are much less significant. Most of the 1D and 2D angular correlation data wherein higher-order harmonic contributions are claimed to exist can be well described without these elements. The transverse momentum ($p_t$) integral ($p_t > 0.15$~GeV/$c$) 2D angular correlations reported for Au+Au collisions at 200 GeV \cite{axialCI} were readily described with physical-model-independent functions consisting of a 2D Gaussian at zero relative opening angle plus the first two terms of a cosine series, the dipole and quadrupole. With this model it was demonstrated that the same-side correlation peak in the data could be easily separated from other structures such as the multipoles via its dependence on relative pseudorapidity and that it could be described accurately with a single 2D Gaussian \cite{axialCI}. This analysis showed that $m > 2$ azimuth harmonics are not {\em required} by these data, although they may be permitted in the model fitting.

Critical evaluation of 2D angular correlation models, including a $m = 3$ cylindrical multipole or sextupole was reported in \cite{axialCI,Tomv3-1,Tomv3-2}. In all three studies the sextupole model element was considered but with different focus. Each study demonstrated that the sextupole was not required by the data.

In Ref.~\cite{axialCI} it was shown that model descriptions of the 2D angular correlations for 62 and 200 GeV Au+Au minimum-bias collisions could achieve a reduction in $\chi^2$ when a sextupole was included. However, the changes in the residuals were not systematically significant given the uncertainties in the data and fitting parameters~\cite{axialCI}.

In all three papers it was shown that the sextupole amplitude was determined by the multipole decomposition of the azimuth projection of the same-side 2D peak, implying that the sextupole derives from the same-side peak rather than from some other aspect of the data. Each analysis demonstrated that the net effect of the additional sextupole in combination with changes in the lower-order dipole and quadrupole terms was equivalent to a same-side 1D peak on azimuth (uniform on relative pseudorapidity), which we refer to as an {\em effective ridge}. Further analysis presented in those papers showed that the effective ridge, in combination with the fitted reduced-amplitude same-side 2D Gaussian, was statistically equivalent to the original same-side 2D Gaussian peak function obtained without the sextupole.  The minor differences were statistically and systematically insignificant but point the way toward explaining why fits including the sextupole achieve reduced $\chi^2$. 

The debate over fitting-model phenomenology may seem unimportant since the entire correlation structure must ultimately be explained by any surviving theory that is not falsified. But in practice the mathematical decomposition of correlation data influences theoretical development. For instance, representing azimuth projections of 2D angular correlations with a single $\cos[2(\phi_1 - \phi_2)]$ term drives and is motivated by a hydrodynamic flow explanation. In fact, any azimuth Fourier component with $m > 1$ is conventionally interpreted as hydrodynamic flow ({\em e.g.} \cite{vmHydro}). The present state of ignorance regarding the physical origin of the same-side 2D peak in more-central heavy-ion collisions does not justify including {\em theoretically-inspired} model elements in the fitting phenomenology which then introduce strong theoretical bias in the data analysis.  The statistical and phenomenological aspects of the fitting model should be tested and well understood first, and then used to guide theoretical development.

The above studies~\cite{axialCI,Tomv3-1,Tomv3-2} address the question of the logical necessity, fitting model instabilities, and net model differences associated with higher-order harmonic terms. In the present analysis the structure of the same-side 2D peak in the angular correlation data for 200 GeV minimum-bias Au+Au collisions~\cite{axialCI} is studied in detail. A single-Gaussian hypothesis is tested using fits to projections of the 2D data onto relative pseudorapidity shown in ~\cite{Tomv3-2}. The net effect of an additional $m=3$ azimuth harmonic is determined and shown to introduce small, non-Gaussian (NG) $\eta$ dependence in the same-side 2D peak. Other NG models are studied, and a single same-side 2D Gaussian hypothesis is tested. The covariation between the amplitudes of the $m=3$ harmonic and the same-side 2D peak is determined and shown to be very large; a same-side invariant combination of model elements is derived. The angular correlation data for the 0-1\% 2.76 TeV Pb+Pb collision data from the ATLAS~\cite{atlas} collaboration, which exhibit an away-side double-peaked structure on azimuth, are described with a variety of fitting models including those with a sextupole.

The paper is organized as follows. In Sec.~\ref{SecII} the correlation measure and standard model function are defined. In Sec.~\ref{SecIII} the single-Gaussian hypothesis is tested for 1D projections of the data. In Sec.~\ref{SecIV} non-Gaussian fitting models are applied to the 2D angular correlation data. Fitting ambiguities and a same-side model invariant are discussed in Sec.~\ref{SecV}. In Sec.~\ref{SecVI} the analysis results for the 0-1\% 2.76 TeV Pb+Pb collision data from ATLAS are presented. A summary and conclusions are given in Sec.~\ref{SecVII}.

\section{Analysis methods}
\label{SecII}

Two-particle correlations from heavy-ion collisions are constructed from charged particle pair histograms on a maximum of six independent momentum coordinates, $p_{t1},\eta_1,\phi_1,p_{t2},\eta_2,\phi_2$. For symmetric collision systems and for measurements near mid-rapidity the correlations on projected 2D sub-spaces $(\eta_1,\eta_2)$ and $(\phi_1,\phi_2)$ are invariant along lines of constant differences $(\eta_1 - \eta_2)$ and $(\phi_1 - \phi_2)$~\cite{aya,axialCI}, thus eliminating two coordinates. The data can be projected onto relative difference coordinates $\eta_\Delta = \eta_1 - \eta_2$ and $\phi_\Delta = \phi_1 - \phi_2$ without loss of information.  Symbols $\Delta\eta$ and $\Delta\phi$ herein refer to intervals in the primary 3D space of the tracking detector and are used to denote the tracking acceptance. The remaining coordinates $p_{t1},p_{t2}$ are integrated over intervals (cuts) that vary with the experiment and specific data set. Angular correlation data are often projected onto 1D variables $\eta_\Delta$ or $\phi_\Delta$. Terms ``near-side'' or ``same-side'' refer to particle pairs with $|\phi_\Delta| \leq \pi/2$ and ``away-side'' refers to $|\phi_\Delta| > \pi/2$. 
 
Measured correlations consist of the pair ratio
\bea
\label{Eq1}
r & \equiv & \rho_{\rm sib} / \rho_{\rm ref}
\eea
which is a ratio of binned quantities corresponding to pairs of particles from the same collision (siblings) and pairs from a reference distribution. In Ref.~\cite{axialCI} the latter was constructed using charged particle pairs from different but similar collision events, or mixed-event pairs denoted by $\rho_{\rm mix}$. Assuming unit-normalized histograms ratio $r$ is approximately 1 such that the difference
\bea
\label{Eq2}
r-1 & = & \frac{\rho_{\rm sib} - \rho_{\rm mix}}{\rho_{\rm mix}}
\equiv \frac{\Delta\rho}{\rho_{\rm mix}}
\eea
carries the significant information.  The above quantity reports the number of correlated pairs per total number of particle pairs on the binned space. Observed correlation structures in relativistic heavy-ion collisions scale roughly in proportion to the event multiplicity with the exception of quantum interference correlations (HBT~\cite{HBT}). The per-pair ratio therefore includes a trivial dependence on inverse multiplicity. Such dependence can be eliminated by using a per-particle measure which is equivalent to Pearson's normalized correlation coefficient~\cite{Pearson} and is defined by
\bea
\label{Eq3}
\frac{\Delta\rho}{\sqrt{\rho_{\rm ref}}} & = & \sqrt{\rho_{\rm ref}}(r-1).
\eea
Prefactor $\sqrt{\rho_{\rm ref}} = d^2N_{\rm ch}/d\eta d\phi$ when the single-particle density is uniform over the acceptance.

Measured $p_t$-integral ($p_t > 0.15$~GeV/$c$) angular correlations from STAR were described with a physical-model-independent fitting function based on examination of the correlation structures resulting from different kinematic, charge-pair combinations, and centrality cuts.  Angular correlations on $(\eta_\Delta,\phi_\Delta)$ for peripheral Au+Au collision data were constructed for pairs with transverse rapidity~\cite{yt} sum $y_{t1} + y_{t2}$ greater than 4 or less than 4 and for like-charge-sign pairs (LS) as well as for unlike-sign pairs (US).

For $y_{t1} + y_{t2} \leq 4$ a 1D peak on $\eta_\Delta$ (and uniform on $\phi_\Delta$) is observed which diminishes with increasing collision centrality~\cite{axialCI,Porter}.  Also at low momentum a sharp peak at $(\eta_\Delta,\phi_\Delta) = (0,0)$ is well represented with a 2D exponential.  Particle identification studies reveal that the sharp peak for US pairs is due to conversion electron backgrounds~\cite{axialCI}. Projections of measured HBT correlations~\cite{STARHBT} onto $(\eta_\Delta,\phi_\Delta)$ show that the LS peak represents quantum correlations. Both are considered backgrounds for this analysis.

Angular correlations for pairs with $y_{t1} + y_{t2} > 4$ reveal a same-side 2D peak at (0,0) plus a broad away-side 1D peak on $\phi_\Delta$ centered at $\phi_\Delta = \pi$ and uniform on $\eta_\Delta$~\cite{axialCI,Porter}.  The latter can be represented by a periodic series of 1D Gaussians on azimuth at odd multiples of $\pi$, or more economically by an away-side dipole $\cos(\phi_\Delta - \pi)$. The dipole representation is accurate for the broad away-side azimuth widths observed in $p_t$-integral angular correlations and for selected $p_t$ intervals up to several GeV/$c$~\cite{DavidHQ}. For higher $p_t$ an away-side Gaussian with reduced width may be more appropriate (see Figs.~10 and 11 in Ref.~\cite{Dihadron1} and also Refs.~\cite{Dihadron2,Dihadron3}). The sum of these four model elements together with a normalization offset (an arbitrary constant determined by the chosen normalization of ratio $r$ above) account for the observed correlations in p+p and peripheral Au+Au collisions from STAR~\cite{aya,axialCI,Porter}.

With increasing centrality this model is insufficient; the residuals indicate the need for additional $\eta_\Delta$-independent azimuth peaks with similar amplitudes centered at $\phi_\Delta = 0$ and $\pi$ which are well represented by a quadrupole term $\cos(2\phi_\Delta)$. No further model elements are required to describe the minimum-bias $p_t$-integral 2D STAR data~\cite{axialCI,etadipole}. Charge-dependent or LS and US correlations as well as correlations obtained with restricted $p_t$ cuts or so-called trigger-associated pair cuts may require different model elements. 

The minimal fitting model for this analysis is therefore given by
\begin{widetext}
\bea
\label{Eq4}
F(\eta_\Delta,\phi_\Delta) & = & A_0 + A_{\rm D} \left[ 1 + \cos(\phi_\Delta - \pi) \right] /2 +
2A_{\rm Q} \cos(2 \phi_\Delta) + A_{\rm 2D} \exp \left\{ - \frac{1}{2} \left[ \left( \frac{\eta_\Delta}{\sigma_{\eta_\Delta}} \right)^2 + \left( \frac{\phi_\Delta}{\sigma_{\phi_\Delta}} \right)^2 \right] \right\} \nonumber \\
 & + & A_{\rm bkg} \exp \left\{ - \left[ \left( \frac{\eta_\Delta}{w_{\eta_\Delta}} \right)^2 + \left( \frac{\phi_\Delta}{w_{\phi_\Delta}} \right)^2 \right]^{\frac{1}{2}} \right\}
  +  A_{\rm soft} \exp \left\{ - \frac{1}{2} \left( \frac{\eta_\Delta}{\sigma_{\rm soft}} 
       \right)^2 \right\} .
\eea
\end{widetext}
This fitting model, which we refer to as the {\em standard model function}, contains 11 free parameters. However, for mid- to most-central Au+Au collisions at RHIC energies the 1D Gaussian ($A_{\rm soft}$ term) is not required. An added sextupole term would have the form $2A_{\rm S} \cos(3 \phi_\Delta)$. Other fitting model elements will be introduced in Secs.~\ref{SecIII} and \ref{SecIV}.

Statistical errors in the fit parameters were calculated using a Taylor-series expansion for the fitting function near the $\chi^2$-minimum. Defining the fit parameters as $\{c_1, c_2, \cdots c_{\rm M} \}$ the above function for arbitrary parameter values near the optimum solution can be expanded as
\bea
\label{Eq5}
F(\eta_\Delta,\phi_\Delta ; c_1 \cdots c_{\rm M}) 
 & = &  F(\eta_\Delta,\phi_\Delta ; c_1^0 \cdots c_{\rm M}^0) \nonumber \\
&  & \hspace{-1.0in} +  \sum_{i=1}^{\rm M} (c_i - c_i^0) \frac{\partial 
F(\eta_\Delta,\phi_\Delta ; c_1 \cdots c_{\rm M})}{\partial c_i}
| _{c_1^0 \cdots c_{\rm M}^0} + \cdots  \nonumber \\
& & \hspace{-1.0in} = F^0(\eta_\Delta,\phi_\Delta) + \sum_{i=1}^{\rm M} \delta c_i
\frac{\partial F}{\partial c_i}|_0 + \cdots ~.
\eea
Superscripts 0 indicate optimum fit values; $\delta c_i = c_i - c_i^0$ and partial derivatives $\partial F/\partial c_i$ are evaluated at the $\chi^2$-minimum. The fit parameter error or covariance matrix $\delta^2 c_{ij}$ is obtained as usual~\cite{FriarNegele} via matrix evaluation
\bea
\label{Eq6}
\delta^2 c_{ij} & = & \delta^2 \left( \delta c_{ij} \right) =
\left[ \sum_{\eta_\Delta,\phi_\Delta} \frac{1}{\epsilon^2(\eta_\Delta,\phi_\Delta)}
\frac{\partial F}{\partial c_i}|_0 \frac{\partial F}{\partial c_j}|_0 \right]^{-1},
\nonumber \\
\eea
where $\epsilon(\eta_\Delta,\phi_\Delta)$ is the bin-wise statistical uncertainty in the correlation data and exponent ($-1$) indicates inversion of the matrix within the square brackets. Diagonal element $\delta^2 c_{ii}$ is the variance $\sigma^2_i$ of the $i^{\rm th}$ parameter. Off-diagonal element $\delta^2 c_{ij}$ is the covariance between the $i^{\rm th}$ and $j^{\rm th}$ parameters also denoted by $cov(i,j)$.

\section{1D projections onto $\eta_\Delta$}
\label{SecIII}

In this section we examine the model-independent form of the $\eta_\Delta$-dependent component of the same-side structure and test its consistency with a 1D Gaussian peak.
We consider 200 GeV Au+Au mid- to most-central collision data~\cite{axialCI} where the same-side 2D peak extends beyond the $\eta$ acceptance and a constant offset plus 1D peak description is possible.

%%%%%%%%%%%%%%%%%%%%%%%%%%%%%%%%%%
\begin{figure}
\includegraphics[keepaspectratio,width=1.75in]{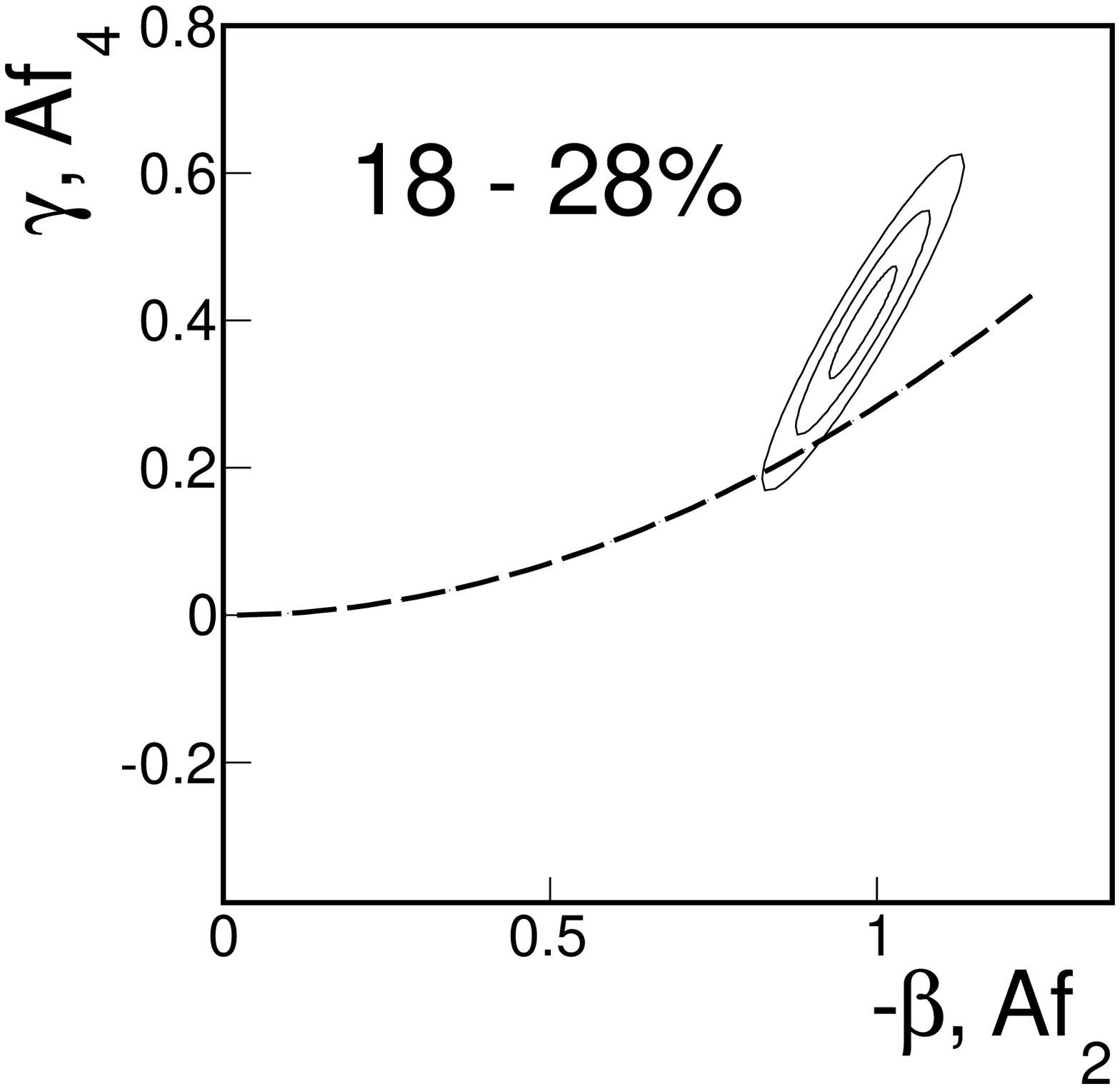}
\put(-95,30){\bf (a)} 
%\put(-95,90){\bf 200 GeV Au+Au}  There is not enough space for this heading
\includegraphics[keepaspectratio,width=1.75in]{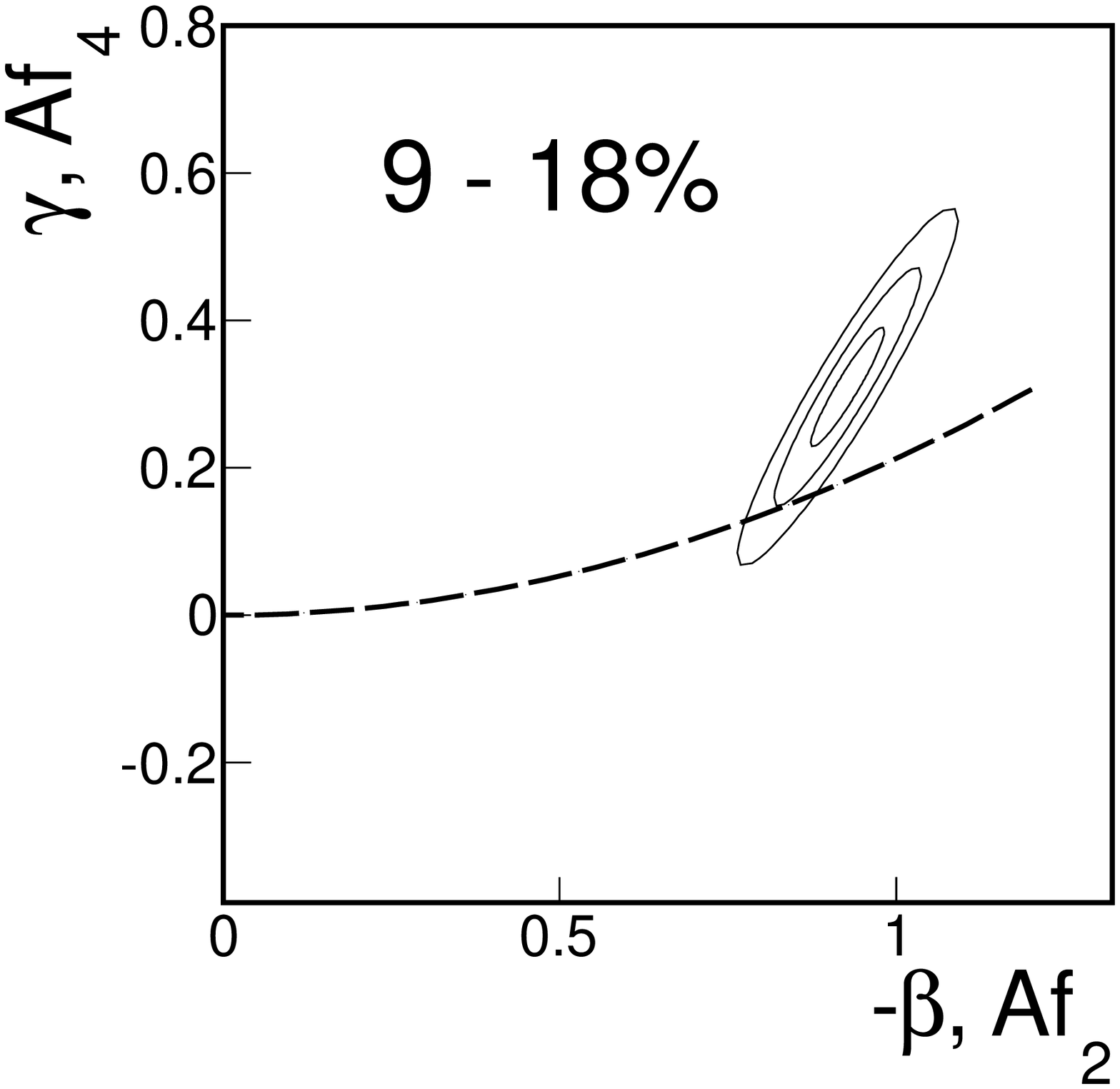}
\put(-95,30){\bf (b)}
\linebreak
\includegraphics[keepaspectratio,width=1.75in]{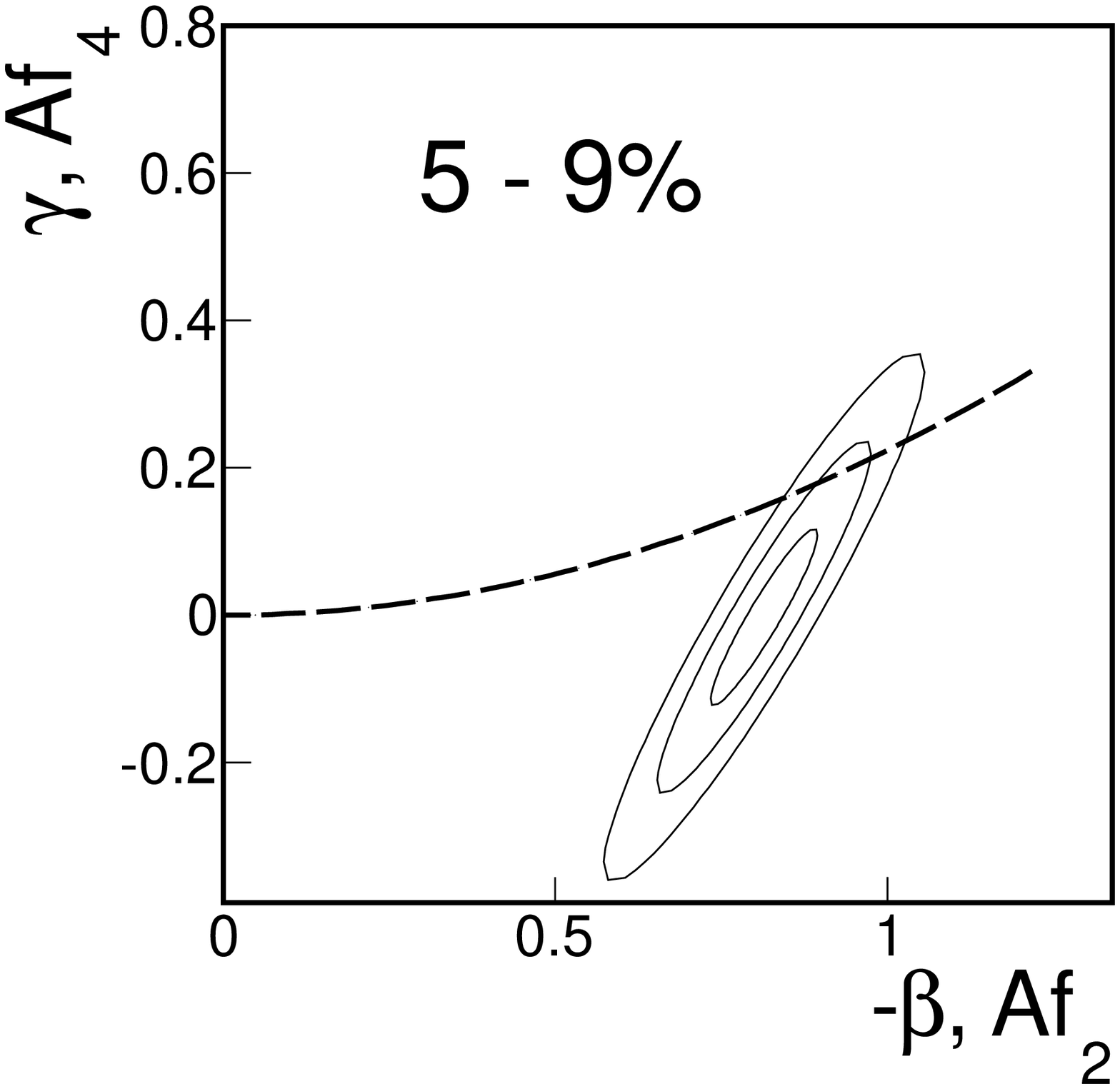}
\put(-95,30){\bf (c)}
\includegraphics[keepaspectratio,width=1.75in]{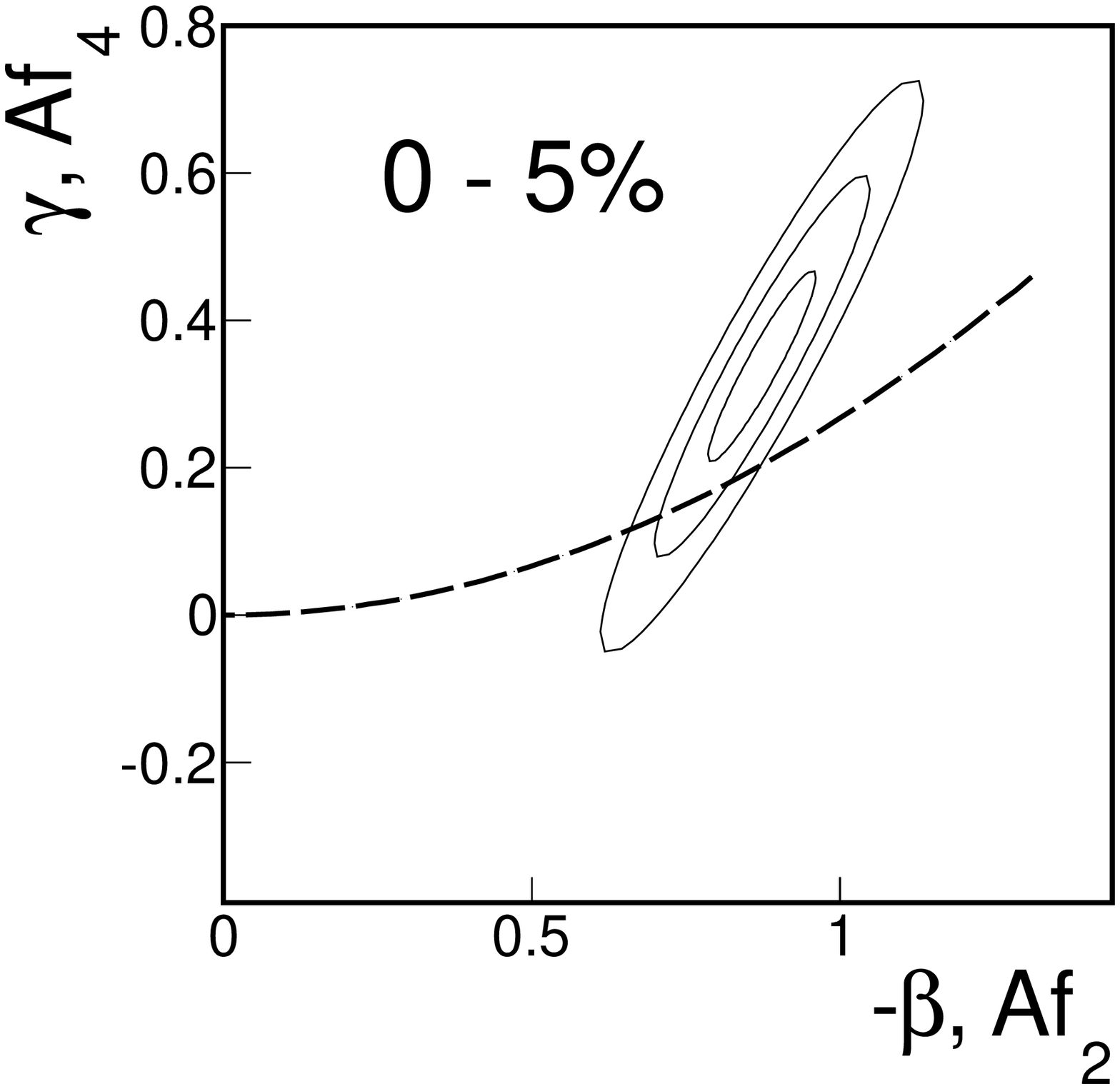}
\put(-95,30){\bf (d)}
\caption{\label{Fig1}
Model independent fitted parameters $(-\beta,\gamma)$ from Eq.~(\ref{Eq7}) with 1, 2 and 3$\sigma$ error contours in comparison with the parabolic {\em locus} of Gaussian model coefficients for the same-side ($|\phi_\Delta| < \pi/2$) projection of the 200 GeV Au+Au correlation data~\cite{axialCI} onto $\eta_\Delta$. The centrality bins shown in panels (a) - (d) correspond to 18-28\%, 9-18\%, 5-9\% and 0-5\%, respectively. The horizontal axes display parameter $-\beta$ from Eq.~(\ref{Eq7}) and $Af_2$ from Eq.~(\ref{Eq9}); the vertical axes represent parameters $\gamma$ and $Af_4$.}
\end{figure}
%%%%%%%%%%%%%%%%%%%%%%%%%%%%%%%%%%

To test the efficacy of the 1D Gaussian description the 2D angular correlations were projected onto $\eta_\Delta$ by averaging the bin values on $\phi_\Delta$ over the interval $[0,\pi/2]$ where the data are periodic and symmetric about $\phi_\Delta$ = 0 by construction. The fitted constant offset ($A_0$), dipole and sharp exponential peak ($A_{\rm bkg}$ element) at (0,0) were subtracted using the values for the standard model function in Eq.~(\ref{Eq4}) and listed in \cite{axialCI}. The quadrupole averages to zero for this selected $\phi_\Delta$ interval. $\eta_\Delta$ projections are shown in \cite{Tomv3-2} for some of these data.

The results were fitted with a model-independent polynomial expansion in $\eta_\Delta$ given by
\bea
\label{Eq7}
F(\eta_\Delta) & = & \alpha + \beta \frac{\eta_\Delta^2}{\Delta\eta^2}
+ \gamma \frac{\eta_\Delta^4}{\Delta\eta^4}
\eea
with free parameters $\alpha,\beta,\gamma$ and STAR TPC tracking acceptance $\Delta\eta$ = 2~\cite{STARTPC}.  Only even powers of $\eta_\Delta$ can contribute due to the symmetry imposed on the 2D histograms.

A Gaussian distribution can be similarly expanded as
\bea
\label{Eq8}
F_{\rm Gauss}(\eta_\Delta) & = & A \exp \left( -\eta_\Delta^2/2\sigma^2_\eta
\right) \nonumber \\
&  & \hspace{-0.25in} \approx  A \left( 1 - \frac{\eta_\Delta^2}{2\sigma^2_\eta}
+ \frac{\eta_\Delta^4}{8\sigma^4_\eta} - \frac{\eta_\Delta^6}{48\sigma^6_\eta}
+ \cdots \right). 
\eea
The magnitudes of the fourth and remaining terms in Eq.~(\ref{Eq8}) are less than the statistical uncertainties in the projected data~\cite{axialCI} for both the full and partial azimuth projections for the more-central collisions corresponding to cross section fractions in the range from 0 to 28\%. The Gaussian hypothesis is therefore tested for the 0-5\%, 5-9\%, 9-18\% and 18-28\% centrality data where the quadratic expansion in $\eta_\Delta^2$ of the Gaussian distribution is statistically accurate. These centralities are most relevant to the sextupole issue owing to the large widths on $\eta_\Delta$. Re-expressing Eq.~(\ref{Eq8}) in the same form as Eq.~(\ref{Eq7}) and truncating at the third term yields
\bea
\label{Eq9} 
 F_{\rm Gauss}(\eta_\Delta) & \approx & A - Af_2 \frac{\eta_\Delta^2}{\Delta\eta^2} + Af_4 \frac{\eta_\Delta^4}{\Delta\eta^4} + \cdots ,
\eea
where $f_2 = \Delta\eta^2/(2\sigma^2_\eta)$ and $f_4 = f_2^2/2$.
The third and remaining terms are all determined by the coefficient of the $\eta_\Delta^2$ term. 

The projected data were fitted with function $F(\eta_\Delta)$ in Eq.~(\ref{Eq7}) where parameter $\alpha$ determines the amplitude $A$ in Eq.~(\ref{Eq9}). The range of possible Gaussian functions is specified by the {\em locus} of ordered pairs $(Af_2,Af_4)$ where $Af_4 = Af_2^2/2$ defines a parabola.

Results are shown in Fig.~\ref{Fig1} for the $[0,\pi/2]$ same-side projection for centrality bins 18-28\%, 9-18\%, 5-9\% and 0-5\% in panels (a) - (d), respectively. One-, two- and three-sigma (3$\sigma$) error contours are shown for fitted parameters $\beta$ and $\gamma$ in Eq.~(\ref{Eq7}) in comparison with the parabolic {\em locus} of Gaussian model coefficients $Af_2$ and $Af_4$.  A Gaussian model description for these 1D data projections lies within 1-3$\sigma$ of the optimum fitted values for all cases. Overall we conclude that a single-Gaussian hypothesis for the 1D $\eta_\Delta$ projected same-side peak is not excluded by the data. However, in all cases $\chi^2$ can be improved by allowing the same-side peak to have a non-Gaussian (NG) geometry, {\em i.e.} $f_4 \neq f_2^2/2$. Non-Gaussian dependence is discussed in the next section.

\section{Non-Gaussian models}
\label{SecIV}

In the previous section it was shown that a single-Gaussian description of the 1D $\eta_\Delta$-projected correlation data was not excluded by the data but that the fit $\chi^2$ could be reduced by allowing minor non-Gaussian dependence in the fitting function. In this section we apply a variety of non-Gaussian fitting models to the entire set of 2D $p_t$-integral angular correlations for 200 GeV minimum-bias Au+Au collisions from STAR~\cite{axialCI}. The study involves the following three steps: (i) We show that the net effect of adding a sextupole term to the standard model function is equivalent to adding a same-side 1D Gaussian peak on azimuth which is uniform on $\eta_\Delta$. Similar results were reported in Refs.~\cite{axialCI,Tomv3-1,Tomv3-2}. (ii) We show that the combination of a same-side 1D Gaussian on azimuth and a reduced 2D Gaussian at (0,0) which reproduces the same-side 2D peak structure in the data is approximately Gaussian with minor non-Gaussian dependence. (iii) We use other ftting models with non-Gaussian forms for the same-side 2D peak and compare the resulting $\chi^2$ values and fit residuals with those produced by a fitting model with an included sextupole element.

\begin{figure*}[t]
\includegraphics[keepaspectratio,width=2.3in]{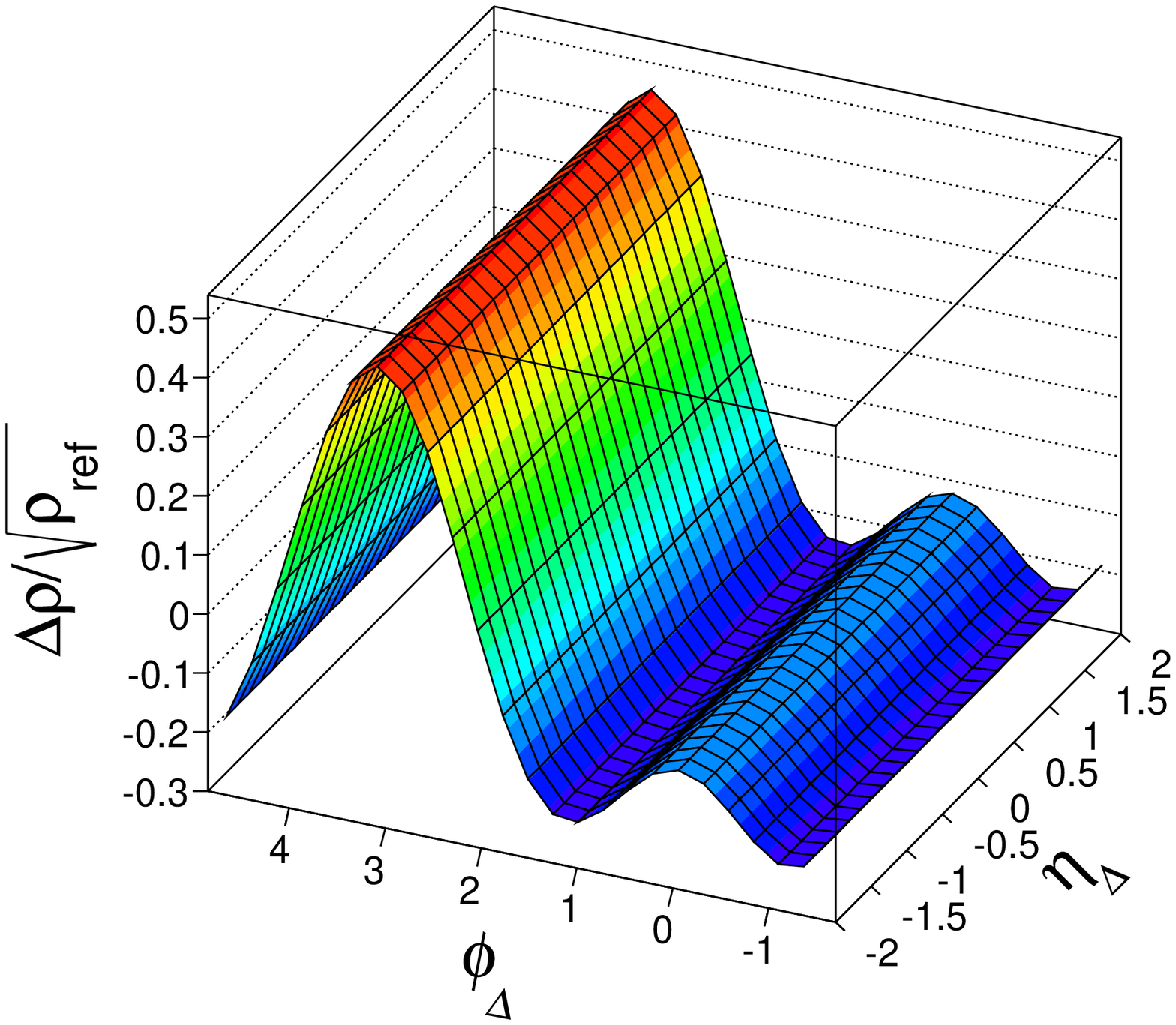}
\put(-60,120){\bf (a)}
\includegraphics[keepaspectratio,width=2.3in]{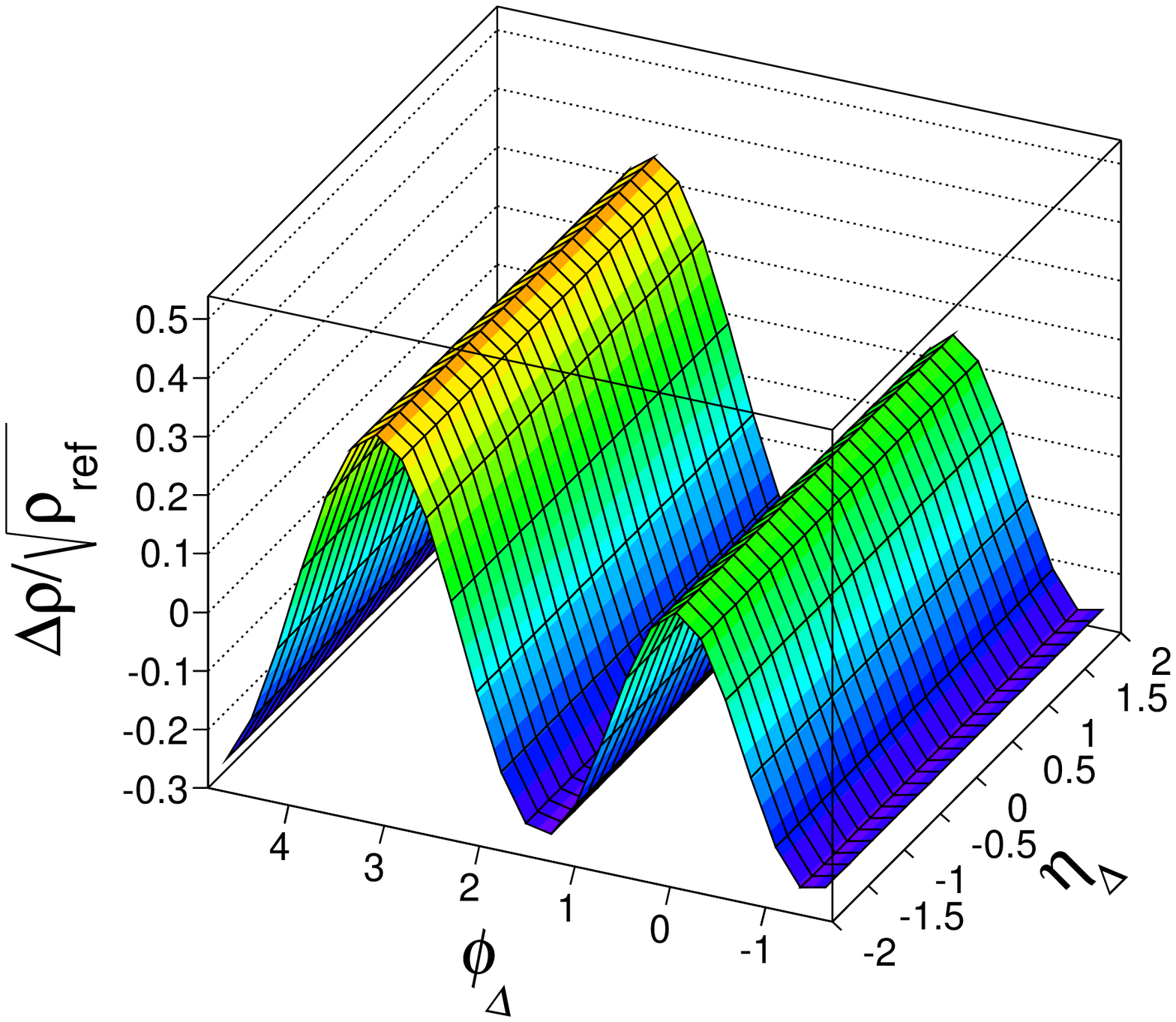}
\put(-60,120){\bf (b)}
\includegraphics[keepaspectratio,width=2.3in]{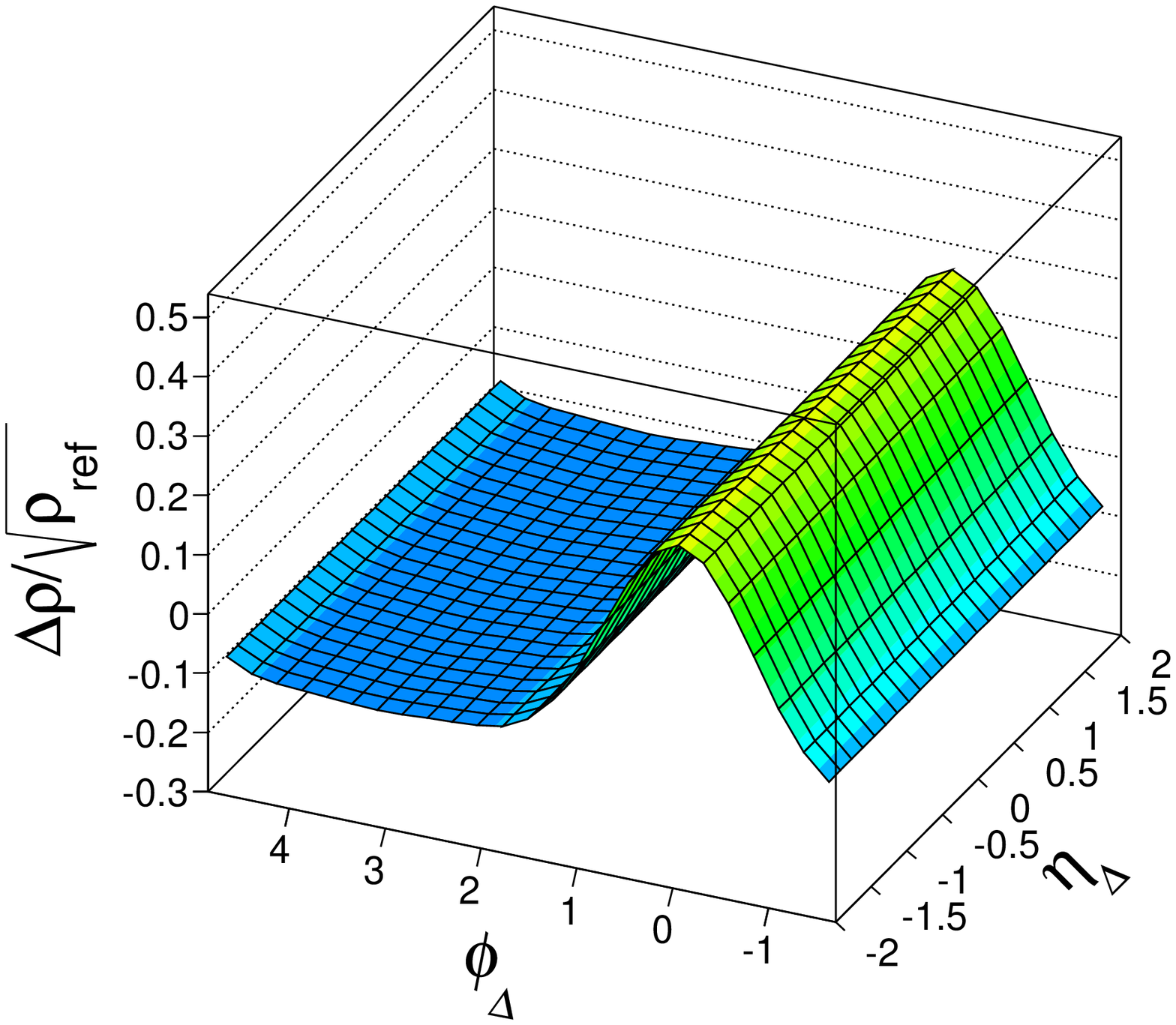}
\put(-60,120){\bf (c)}
\caption{\label{Fig3}
(Color online) Panel (a): fitted dipole + quadrupole. Panel (b): multipole sum including a sextupole. Panel (c):  difference (b) - (a). The multipoles are from fits to the 200 GeV Au+Au 9-18\% centrality data from STAR~\cite{axialCI} using the standard model function in Eq.~(\ref{Eq4}) plus a sextupole term.}
\end{figure*}

The standard model function in Eq.~(\ref{Eq4}) describes the $\eta_\Delta$-independent structure in the away-side correlation data~\cite{aya,axialCI}. Adding a third harmonic model element (sextupole) forces the dipole and quadrupole terms to adjust in order to maintain a good fit to the away-side data. The net difference between the sum of multipoles obtained by fitting with and without the sextupole term is shown in Fig.~\ref{Fig3} for the 9-18\% centrality bin. The right-most panel shows the quantity [$A_{\rm D}^{\prime}\cos(\phi_\Delta - \pi)/2 + 2A_{\rm Q}^{\prime}\cos(2\phi_\Delta) + 2A_{\rm S}^{\prime}\cos(3\phi_\Delta)$ $-$ $A_{\rm D}\cos(\phi_\Delta - \pi)/2 - 2A_{\rm Q}\cos(2\phi_\Delta)$] where primes indicate fitting parameters obtained with an included sextupole.  The net structural difference is a narrow, same-side 1D peak (effective ridge) on azimuth as previously demonstrated in Refs.~\cite{axialCI,Tomv3-1,Tomv3-2}.

%%%%%%%%%%%%%%%%%%%%%%%%%%%%%%%%%%
%%%%%%%%%%%%%%%%%%%%%%%%%%%%%%%%%%

This effective ridge can be accurately represented as a periodic 1D Gaussian, where
\bea
\label{Eq10}
A_{\rm SSG}\sum_{k=0,\pm2,\cdots} e^{-(\phi_\Delta - k\pi)^2/2\sigma^2_{\rm S}}
  =  \sum_{m \geq 0} a_m \cos(m\phi_\Delta)  \nonumber \\
  =  \frac{A_{\rm SSG}\sigma_{\rm S}}{2\pi} \left[ 1 + 2\sum_{m=1}^{\infty} e^{-m^2 \sigma^2_{\rm S} /2}
\cos(m\phi_\Delta) \right],
\eea
and subscript ``SSG'' means same-side Gaussian.
The multipole summation rapidly converges for typical widths $\sigma_{\rm S}$ of order 0.7 such that amplitudes for $m > 3$ are negligible.  Adding a same-side 1D azimuth Gaussian to the standard model function in Eq.~(\ref{Eq4}) with $\sigma_{\rm S}$ of order 0.7 is statistically equivalent to adding same-side dipole, quadrupole and same-side sextupole elements where the amplitude of the sextupole element is
\bea
\label{Eq11}
A_{\rm S} & = & \frac{A_{\rm SSG}\sigma_{\rm S}}{2\pi} e^{-9\sigma^2_{\rm S}/2}.
\eea
Therefore, including a sextupole in the fitting model reduces the away-side dipole, increases the quadrupole and reduces the amplitude and $\eta_\Delta$ width of the same-side 2D peak function.

Fitting the data with an added sextupole element is statistically equivalent to fitting the data with an additional 1D same-side azimuth Gaussian whose width is approximately 0.7.  Given the amplitude of the sextupole it is straightforward to calculate the amplitude of the same-side 1D Gaussian equivalent and vice-versa. The exponential factor in Eq.~(\ref{Eq11}) $\exp(-9\sigma^2_{\rm S}/2)$ accounts for the fact that a relatively large ($\approx$ 50\% of $A_{\rm 2D}$) same-side 1D Gaussian equivalent amplitude corresponds to a relatively small (few percent of $A_{\rm 2D}$) sextupole amplitude. For example, if $\sigma_{\rm S}$ = 0.7 the amplitude of the same-side Gaussian equivalent is $A_{\rm S}/0.012$, a large amplification factor. Increased magnitude of $A_{\rm SSG}$ directly reduces $A_{\rm 2D}$ in the fitting. 

%%%%%%%%%%%%%%%%%%%%%%%%%%%%%%%%%%
\begin{figure*}[t]
\includegraphics[keepaspectratio,width=2.3in]{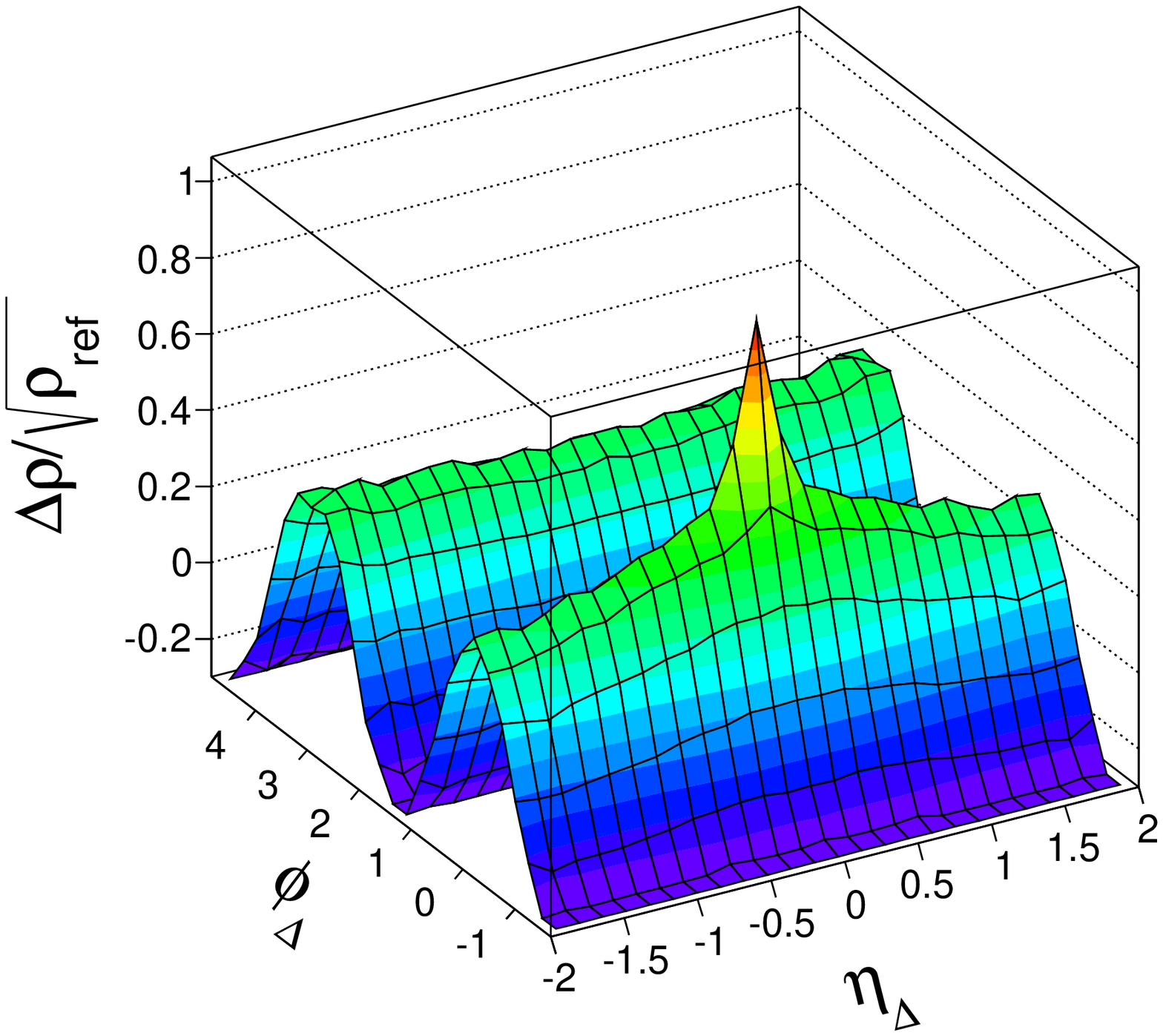}
\put(-125,115){\bf (a)}
\includegraphics[keepaspectratio,width=2.3in]{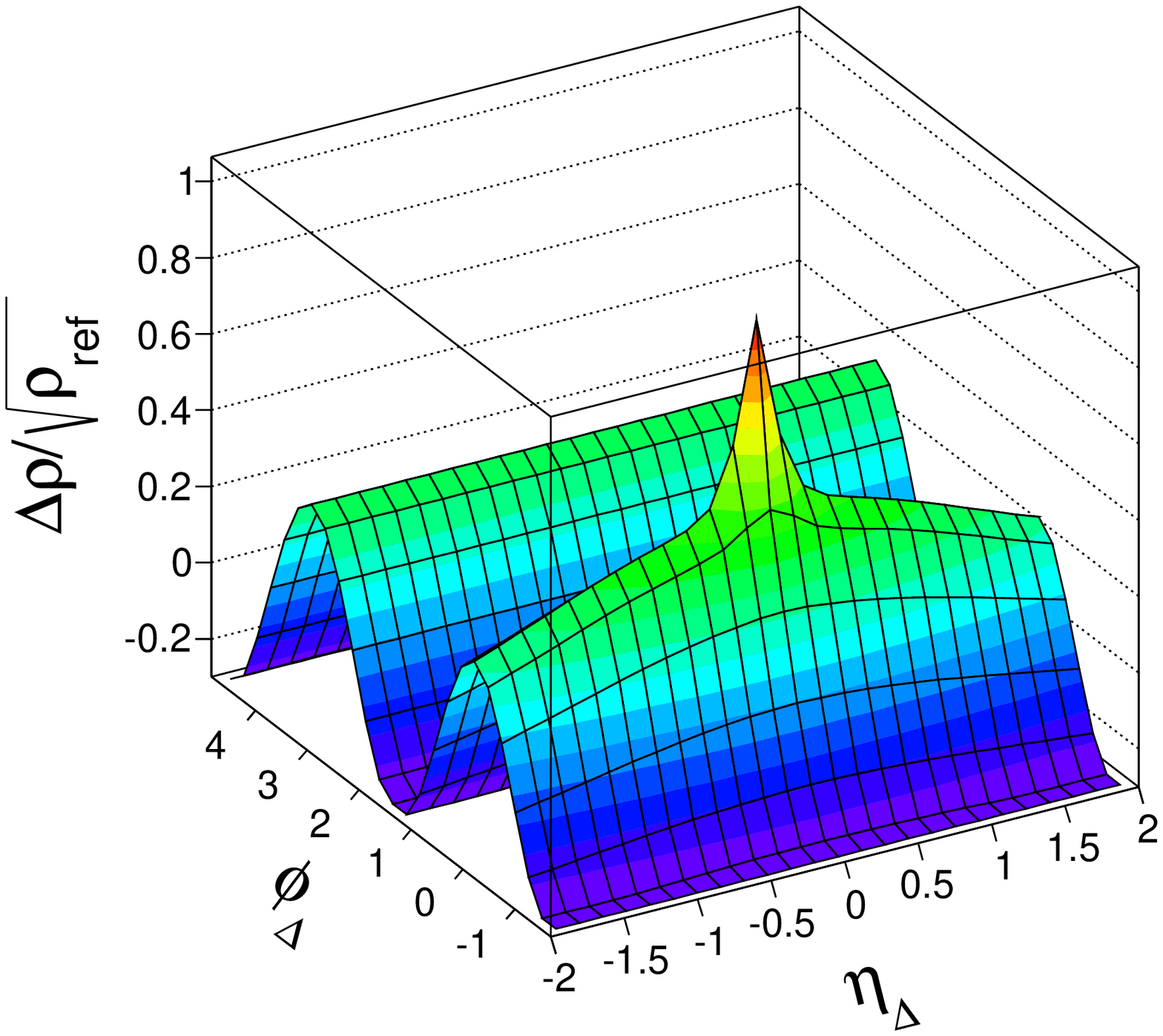}
\put(-125,115){\bf (b)}
\includegraphics[keepaspectratio,width=2.3in]{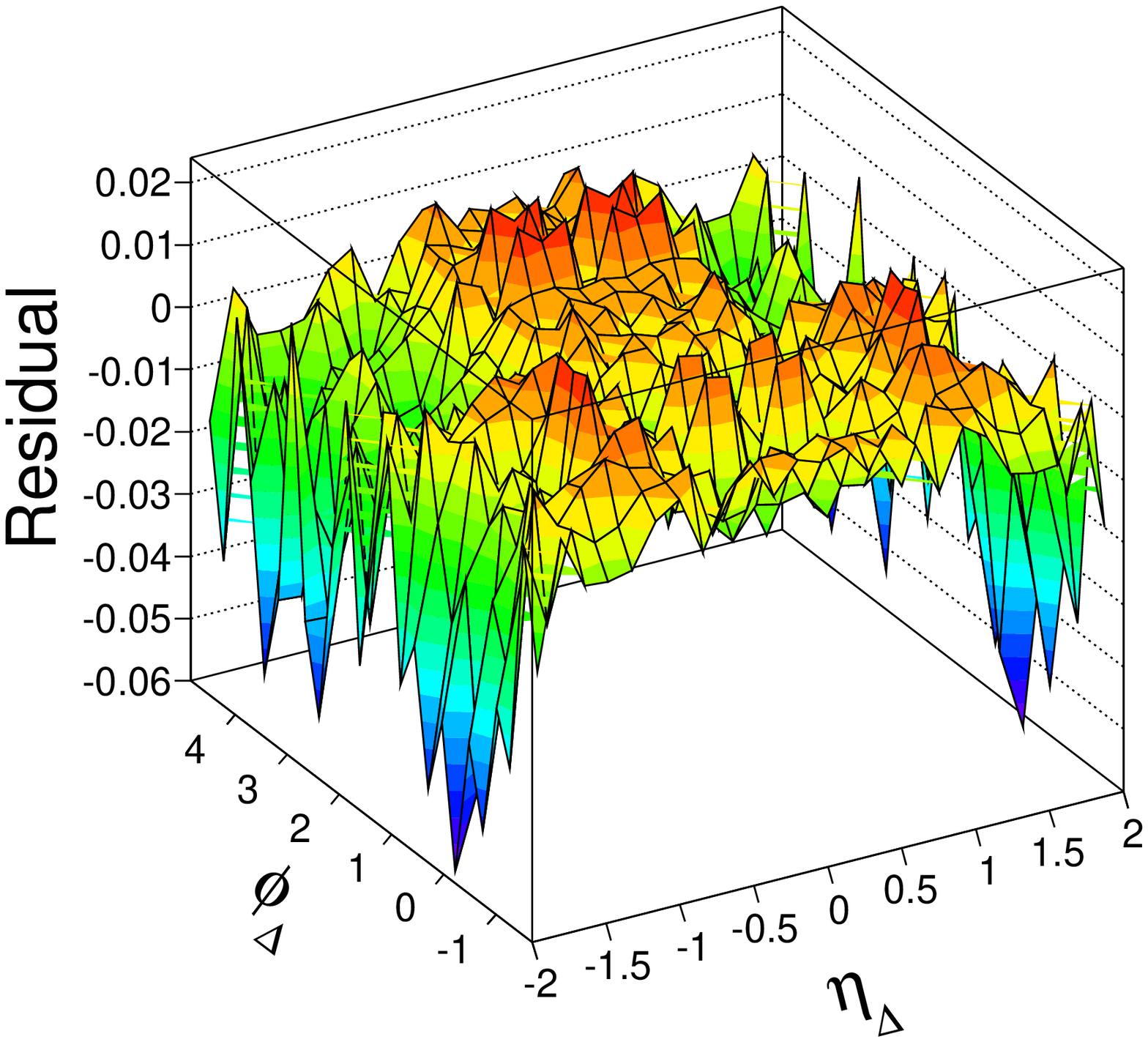}
\put(-125,115){\bf (c)}
\linebreak
\includegraphics[keepaspectratio,width=2.3in]{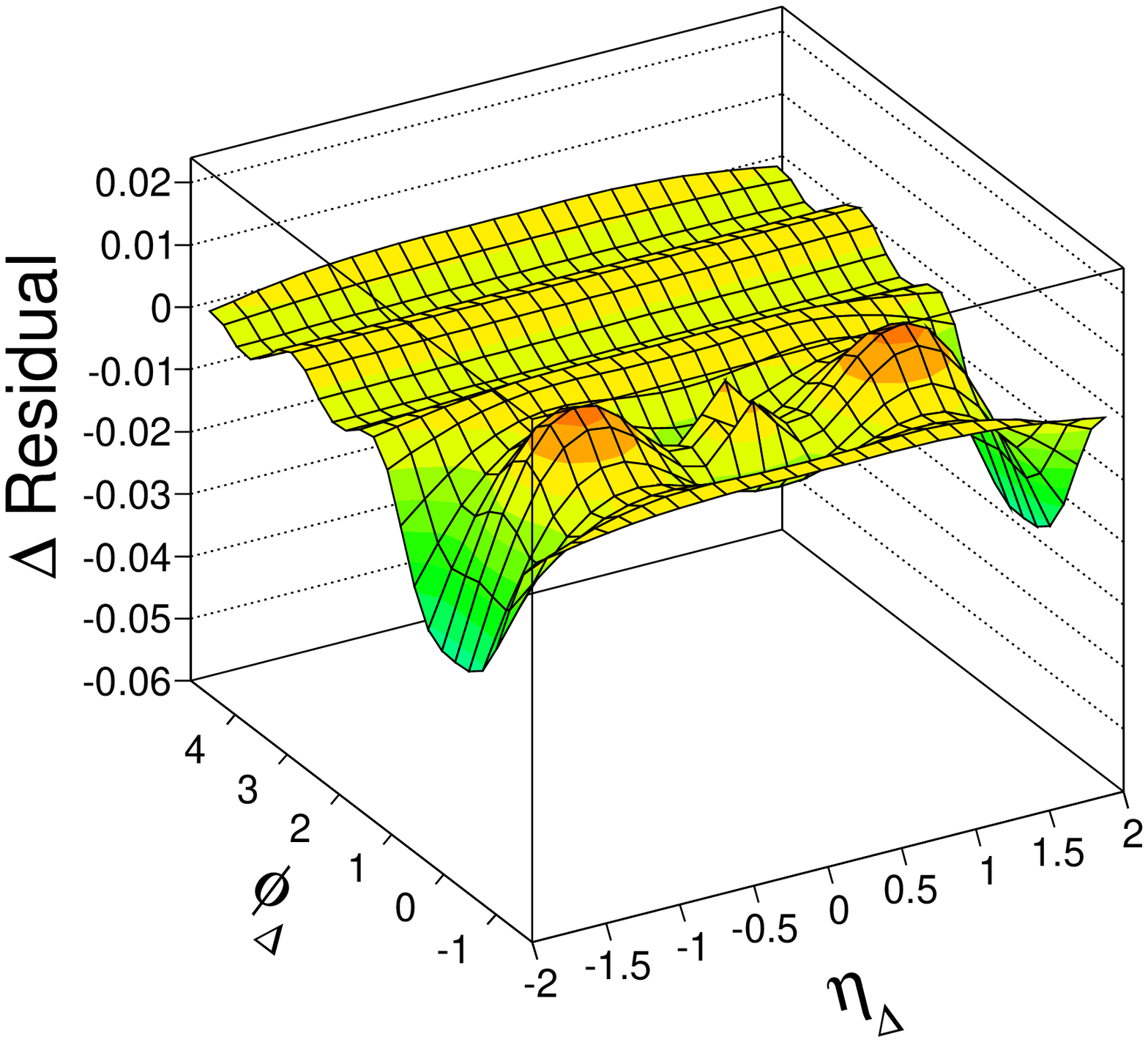}
\put(-125,115){\bf (d)}
\includegraphics[keepaspectratio,width=2.3in]{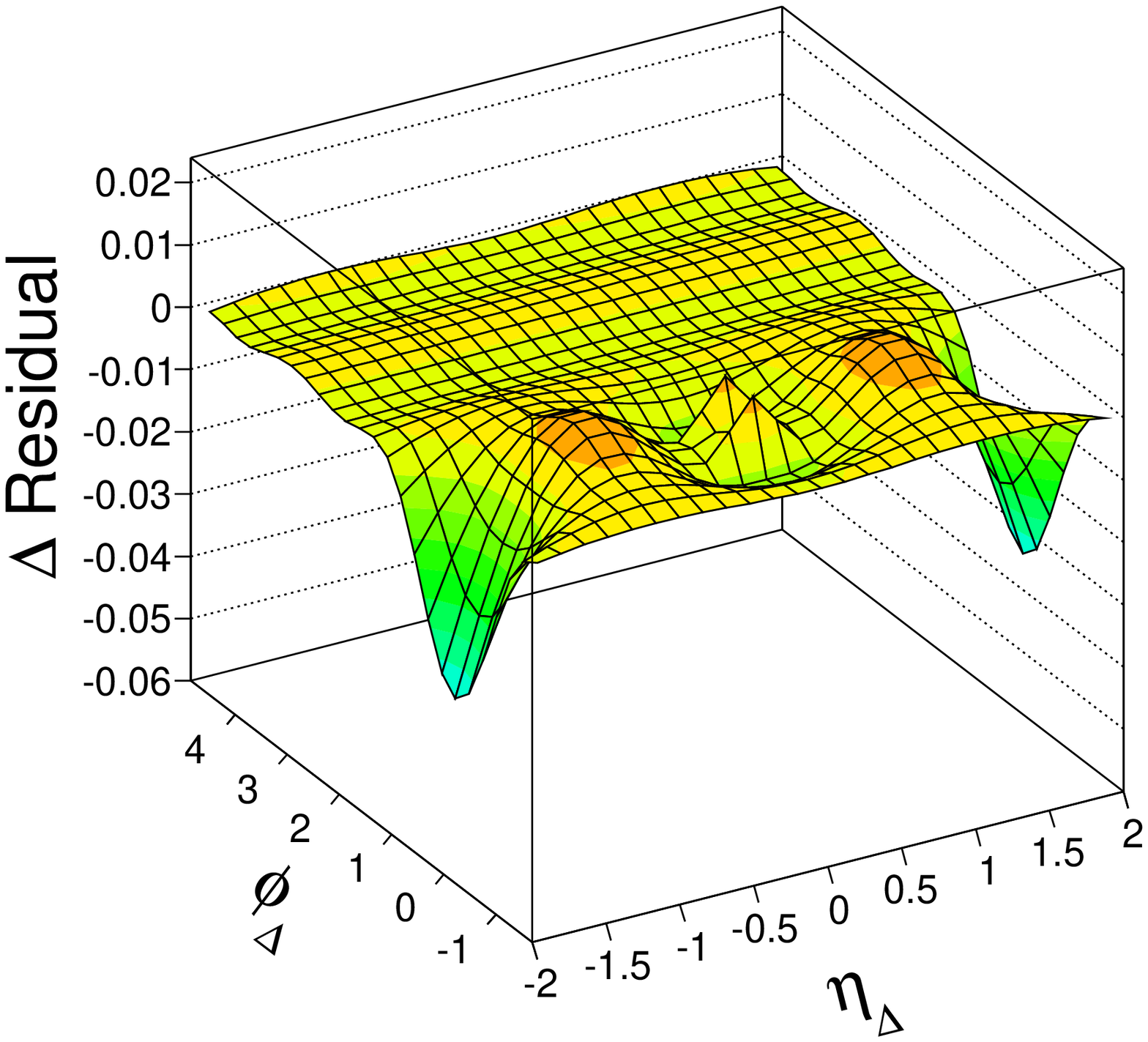}
\put(-125,115){\bf (e)}
\includegraphics[keepaspectratio,width=2.3in]{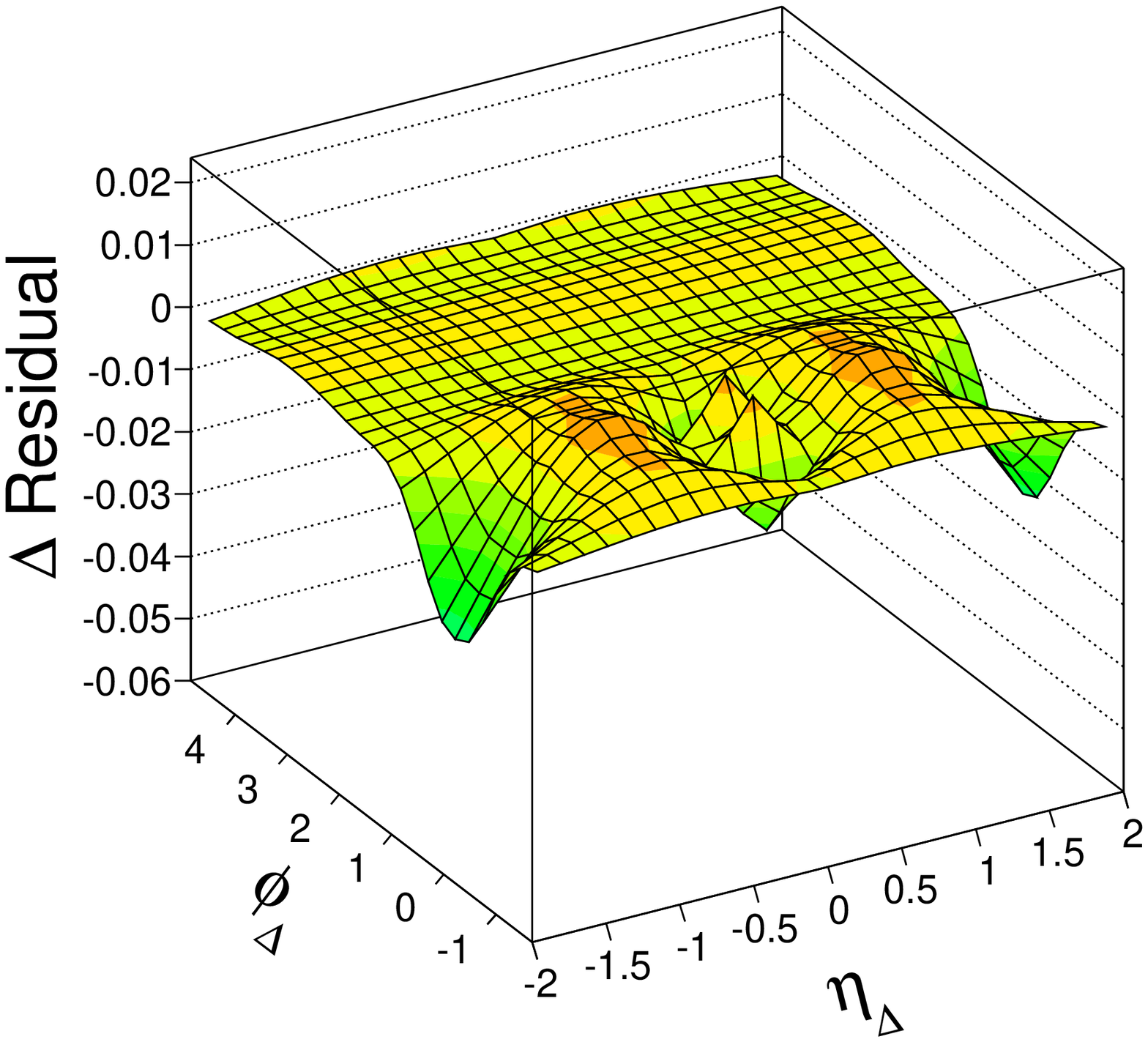}
\put(-125,115){\bf (f)}
\linebreak
\includegraphics[keepaspectratio,width=2.3in]{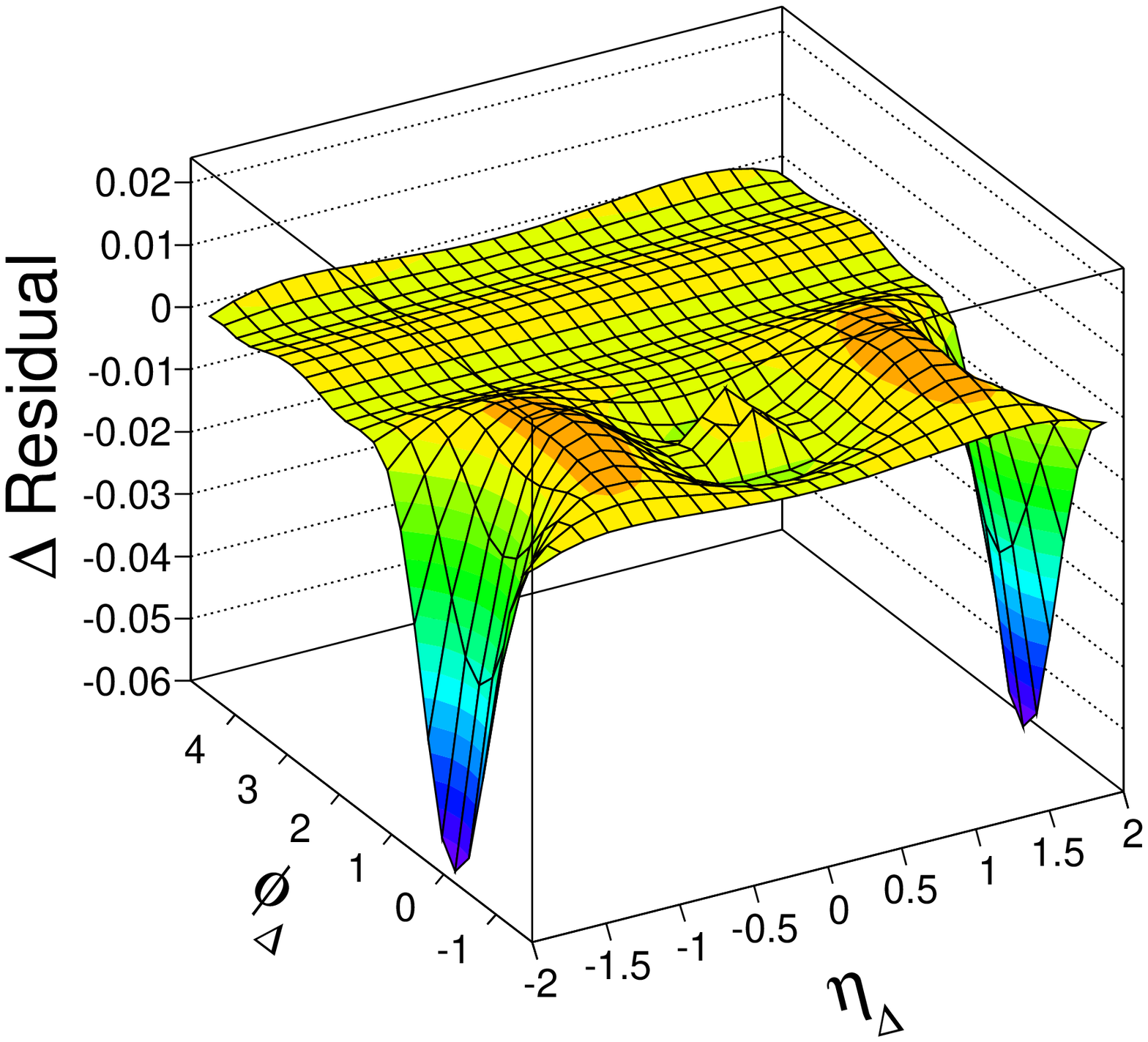}
\put(-125,115){\bf (g)}
\includegraphics[keepaspectratio,width=2.3in]{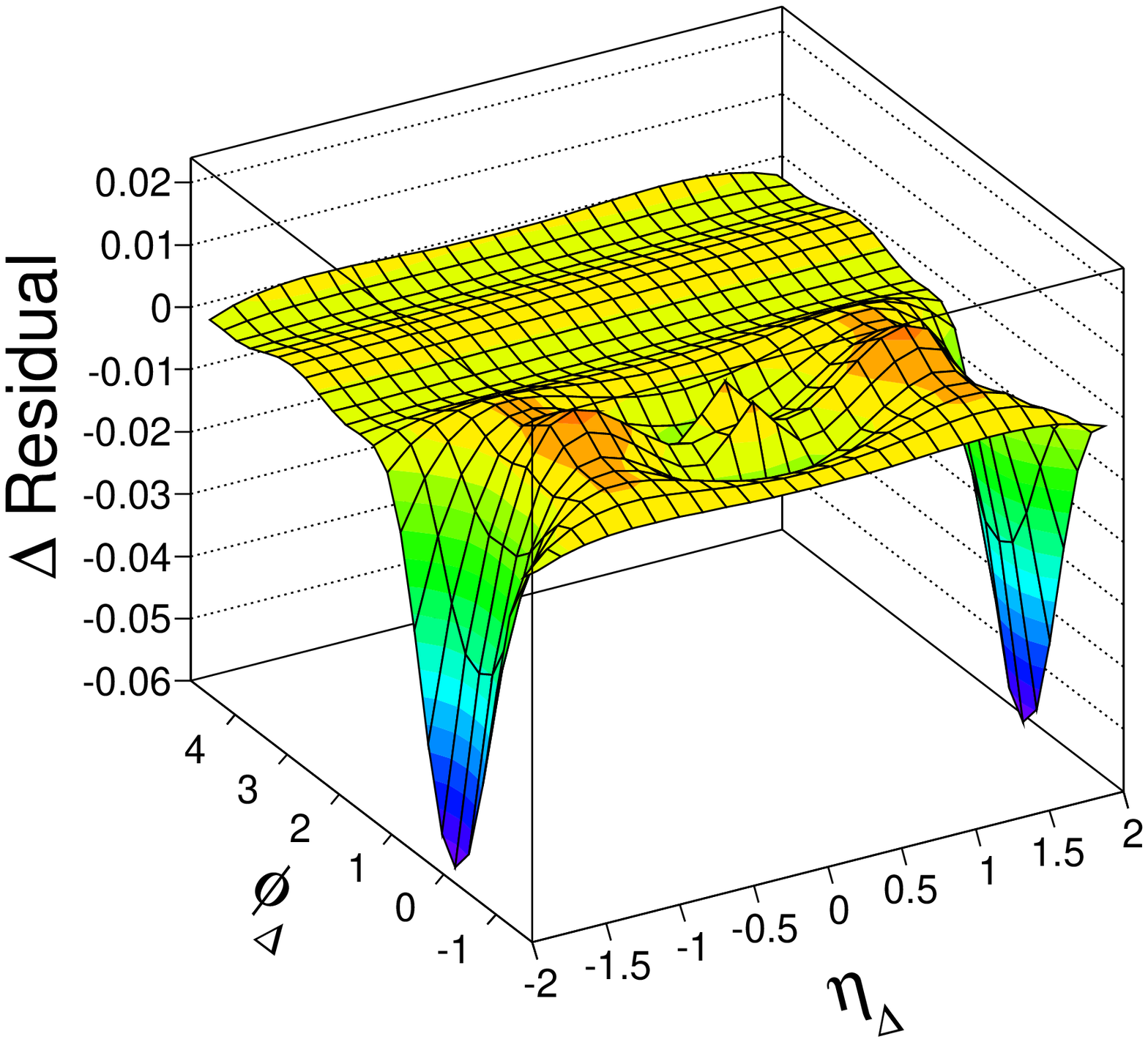}
\put(-125,115){\bf (h)}
\includegraphics[keepaspectratio,width=2.3in]{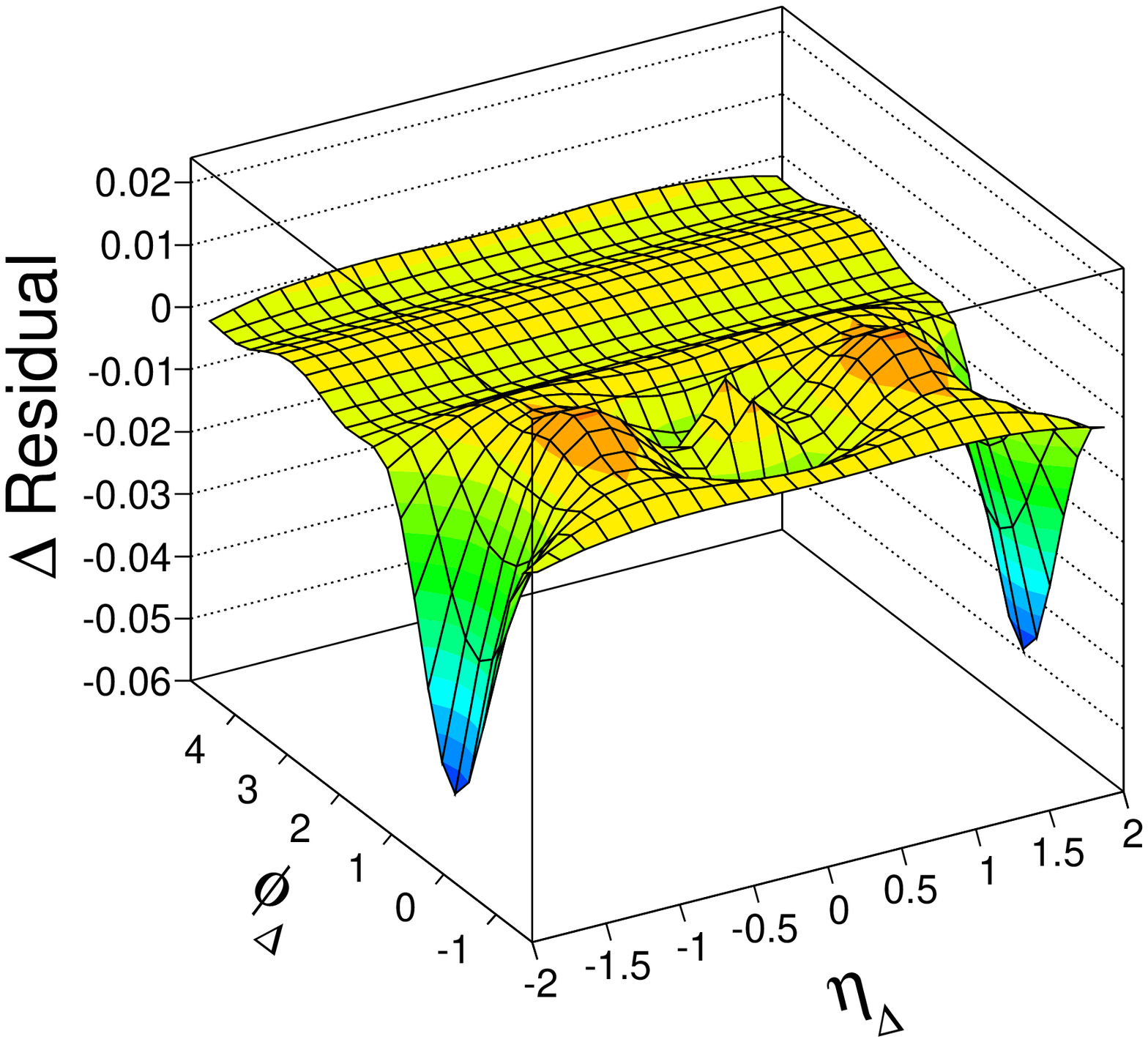}
\put(-125,115){\bf (i)}
\caption{\label{Fig6}
(Color online) Panel (a): angular correlations from 200 GeV Au+Au collision data~\cite{axialCI} for the 28-38\% centrality bin. Panel (b): best fit standard model function. Panel (c): residuals = standard model function $-$ data. Panels (d) - (i) show the differences (standard model function residuals $-$ NG model residuals) for the sextupole, SSG, NG exponents, $\eta_\Delta$ polynomial, $\eta_\Delta$ polynomial with NG $\phi_\Delta$ exponent, and quartic models, respectively. In panels (c) - (i) the vertical scales are all increased by 16$\times$.}
\end{figure*}
%%%%%%%%%%%%%%%%%%%%%%%%%%%%%%%%%%

%%%%%%%%%%%%%%%%%%%%%%%%%%%%%%%%%%
\begin{figure*}
\includegraphics[keepaspectratio,width=3.4in]{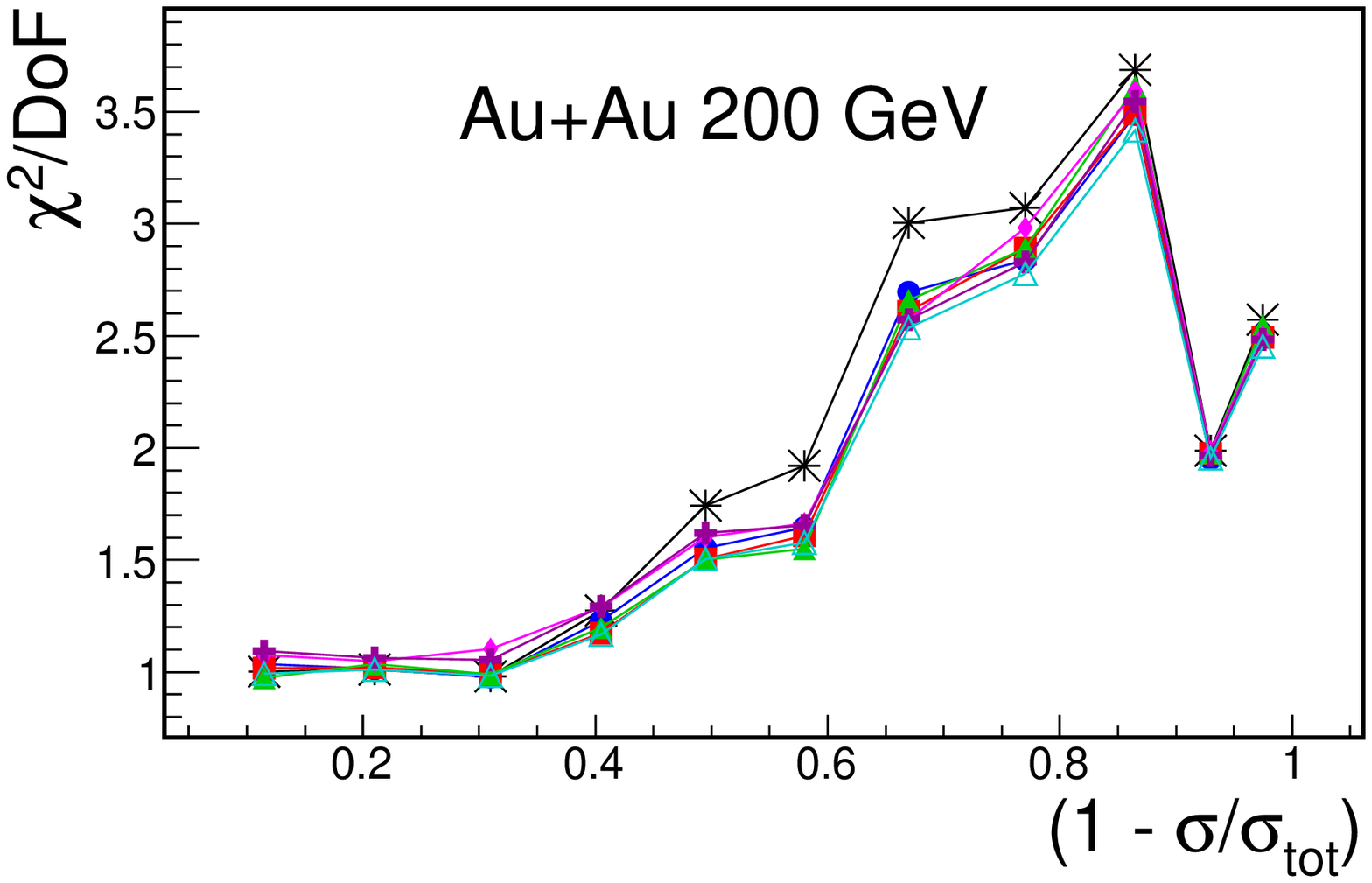}
\put(-200,120){\bf (a)}
\includegraphics[keepaspectratio,width=3.4in]{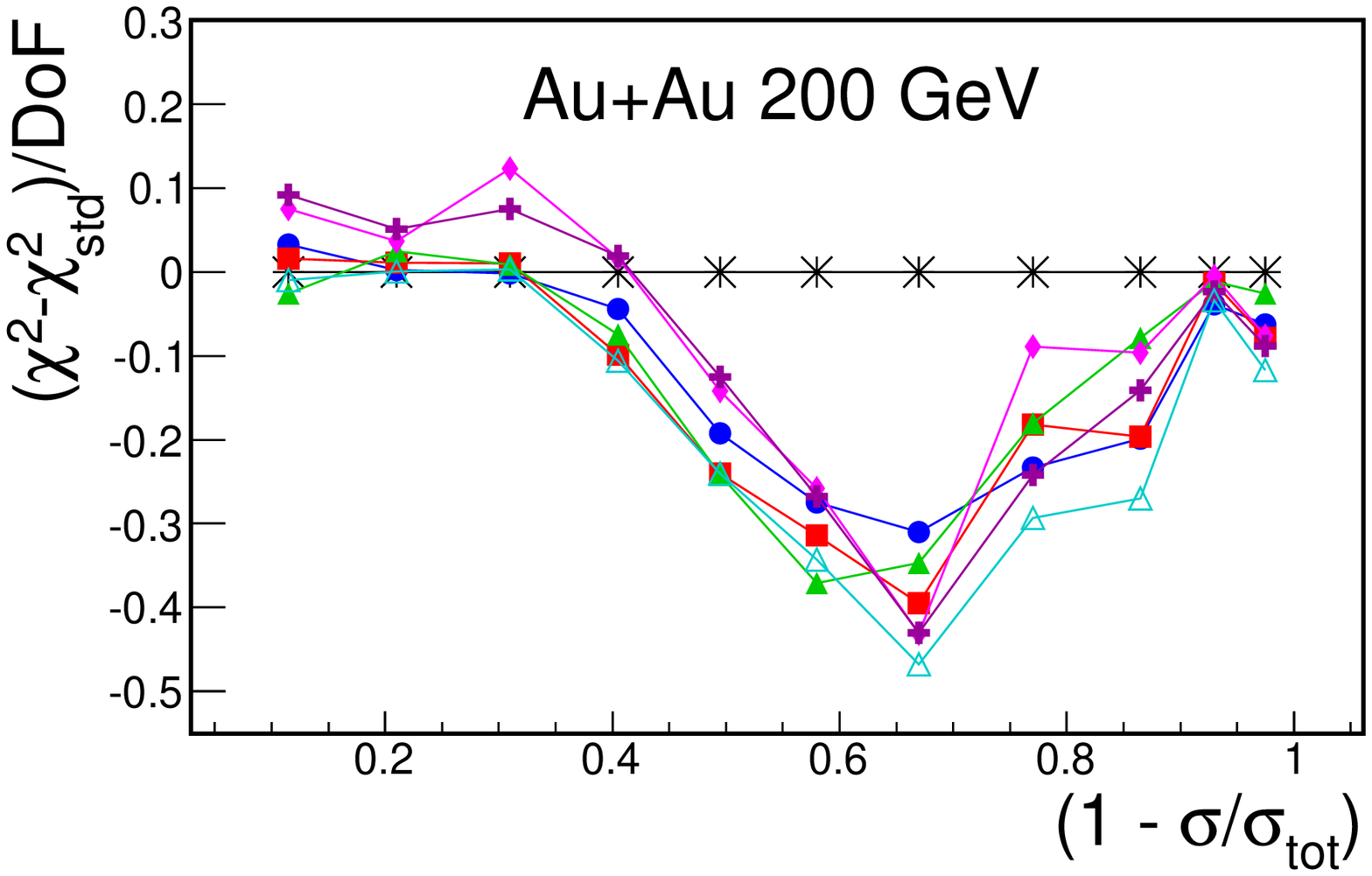}
\put(-200,120){\bf (b)}
\caption{\label{Fig5}
(Color online) Panel (a): $\chi^2$/DoF versus collision centrality for standard and non-Gaussian model functions discussed in the text (symbols). Panel (b): differences between the $\chi^2$/DoF for the six non-Gaussian models and that for the standard model function (symbols) versus centrality. Lines connect corresponding model results. The symbols denote each model as follows: (1) black stars, standard, Eq.~(\ref{Eq4}); (2) solid blue circles, sextupole, Eq.~(\ref{Eq11}); (3) solid red squares, SSG, Eq.~(\ref{Eq10}); (4) solid green triangles, NG exponents, Eq.~(\ref{Eq13}); (5) solid magenta diamonds, $\eta_\Delta$ polynomial, Eq.~(\ref{Eq14}); (6) solid brown ``plus'' symbols, $\eta_\Delta$ polynomial with NG $\phi_\Delta$ exponent, Eq.~(\ref{Eq15}); (7) open cyan triangles, quartic, Eq.~(\ref{Eq16}).}
\end{figure*}
%%%%%%%%%%%%%%%%%%%%%%%%%%%%%%%%%%

Refering to Sec.~\ref{SecIII} and using Eqs.~(\ref{Eq8}) and (\ref{Eq9}) it is easy to demonstrate that adding a constant $B$ to a 1D Gaussian on $\eta_\Delta$ changes the relation between the coefficients of the $\eta_\Delta^2$ and $\eta_\Delta^4$ terms given by
\bea
\label{Eq12}
F_{\rm Gauss}(\eta_\Delta) & \rightarrow & A\exp \left( -\eta_\Delta^2/2\sigma_\eta^2 \right) + B \nonumber \\
 & & \hspace{-0.75in} \approx  (A+B) \left(1 - f_2^{\prime} \frac{\eta_\Delta^2}{\Delta\eta^2}
 + f_4^{\prime} \frac{\eta_\Delta^4}{\Delta\eta^4} + \cdots \right),
\eea
where $f_2^{\prime} = Af_2/(A+B)$ and $f_4^{\prime} = Af_4/(A+B)$ ($f_2$ and $f_4$ are defined in Sec.~\ref{SecIII}). In Eq.~(\ref{Eq12}) $f_4^{\prime} \neq f_2^{\prime 2}/2$, meaning that the combination is non-Gaussian. Fig.~\ref{Fig1} shows that minor deviations from a pure Gaussian function reduce the $\chi^2$ for fits to the 1D $\eta_\Delta$ projections of the same-side data. Adding a sextupole term to the standard model function in Eq.~(\ref{Eq4}) is equivalent to adding a same-side 1D Gaussian. $\chi^2$ is reduced when a sextupole is included in the fitting because it introduces relatively minor non-Gaussian dependence in the model description of the same-side 2D peak.

Given this insight it is interesting to explore other non-Gaussian models for the same-side 2D peak distribution. Several examples are provided here to illustrate the range of possible effects on $\chi^2$ and the fit residuals, but the list of examples is not exhaustive~\cite{etadipole}. The non-Gaussian models include one in which a same-side 1D azimuth Gaussian [as in Eq.~(\ref{Eq10})] is added to Eq.~(\ref{Eq4}) and four in which the 2D peak-model element in Eq.~(\ref{Eq4}) is modified.

The non-Gaussian modifications are as follows: (i) Replace both exponents with fit parameters $\gamma$ and $\delta$. The $A_{\rm 2D}$ term then becomes
\bea
\label{Eq13}
A_{\rm 2D} \exp \left\{ - \frac{1}{2} \left[ \left( \frac{\eta_\Delta}{\sigma_{\eta_\Delta}} \right)^{2\gamma} + \left( \frac{\phi_\Delta}{\sigma_{\phi_\Delta}} \right)^{2\delta} \right] \right\},
\eea
where $\gamma \neq 1$ and/or $\delta \neq 1$ produce a non-Gaussian 2D function, the product of two factors. The NG model in the $\eta_\Delta$ factor is expected to be more significant than that in the $\phi_\Delta$ factor. The NG azimuth dependence is introduced to confirm that such NG effects are negligible. (ii) Replace the $\eta_\Delta$-dependent factor by a power series resulting in 2D peak model
\bea
\label{Eq14}
A_{\rm 2D} \left( 1 + \alpha \frac{\eta_\Delta^2}{\Delta\eta^2}
+ \beta \frac{\eta_\Delta^4}{\Delta\eta^4} \right)
\exp \left[ - \frac{1}{2} \left( \frac{\phi_\Delta}{\sigma_{\phi_\Delta}} \right)^2 \right]
\eea
with additional fitting parameters $\alpha$ and $\beta$. (iii) Use a form similar to the preceding in which the $\phi_\Delta$-dependent exponent is allowed to vary
\bea
\label{Eq15}
A_{\rm 2D} \left( 1 + \alpha \frac{\eta_\Delta^2}{\Delta\eta^2}
+ \beta \frac{\eta_\Delta^4}{\Delta\eta^4} \right)
\exp \left[ - \frac{1}{2} \left( \frac{\phi_\Delta}{\sigma_{\phi_\Delta}} \right)^{2\delta} \right].
\eea
(iv) Add quartic $\eta_\Delta^4$ and $\phi_\Delta^4$ terms in the exponential arguments
\bea
\label{Eq16}
A_{\rm 2D} \exp \left\{ - \frac{1}{2} \left[ \left( \frac{\eta_\Delta}{\sigma_{\eta_\Delta}} \right)^2 + \lambda \eta_\Delta^4 +  \left( \frac{\phi_\Delta}{\sigma_{\phi_\Delta}} \right)^2 + \zeta \phi_\Delta^4 \right] \right\},
\nonumber \\
\eea
where $\lambda$ and $\zeta$ are additional fitting parameters. The factorized form of the 2D peak function is maintained in each case. The sextupole term was excluded from fits with NG functions in Eqs.~(\ref{Eq13}) - (\ref{Eq16}).

These five non-Gaussian fitting models plus the standard model function with and without the sextupole were used to fit all of the angular correlation data for 200 GeV minimum-bias Au+Au collisions from STAR~\cite{axialCI}. The largest overall reduction in $\chi^2$/DoF occurs for the 28-38\% centrality data. For each fitting model the residuals (model $-$ data) are very similar and are very small relative to the principal correlation structures in the data.

In Fig.~\ref{Fig6} the data, fitted standard model function and residuals (for which the scale has been expanded by 16$\times$) for the 28-38\% centrality data are shown in the upper row of panels. The small differences between the residuals for the standard model function and those for the six non-Gaussian models (residuals for the standard model function $-$ residuals for the NG models) are shown in the remaining panels of Fig.~\ref{Fig6} using the same expanded scale. The differences between residuals are equivalent to the differences between the model functions. None of the present fitting models addresses the away-side residuals~\cite{etadipole}. From the results in panels (d) - (i) it is clear that the important non-Gaussian dependence explored here is on $\eta_\Delta$, while any possible non-Gaussian dependence on $\phi_\Delta$ is insignificant; the Gaussian $\phi_\Delta$ factor in Eq.~(\ref{Eq4}) ($\delta$ = 1) suffices in all cases.

The principal features in the residuals for the standard model function which indicate non-Gaussian dependence in the same-side correlation structure are the small peaks near $|\eta_\Delta|$ = 1 and the dips at $|\eta_\Delta|$ = 0.0-0.5 and 2. Those features are approximately 3$\sigma$, 2$\sigma$ and 2$\sigma$, respectively, relative to the statistical uncertainties in the data and are consistent with a same-side peak on $\eta_\Delta$ that is leptokurtic relative to a Gaussian. The larger-magnitude dips at the acceptance edges are especially indicative of long-range non-Gaussian tails in the data distribution. Each of the six non-Gaussian models (including the sextupole model element) reduces the same-side residuals by similar amounts.

The systematic uncertainties in the correlation data and in the parameter values of the standard model function are comparable to or greater than the corresponding statistical uncertainties~\cite{axialCI}. Therefore, the systematic significance of the residuals of the standard model function and the non-Gaussian dependence of the $\eta_\Delta$ factor of the same-side 2D peak is reduced. A 2D-Gaussian hypothesis is not excluded by these data. This result is consistent with the conclusions of the model-independent analysis of the same-side 1D $\eta_\Delta$ projections presented in Sec.~\ref{SecIII}.

The best-fit values of $\chi^2$ per degree-of-freedom (DoF) for all models and collision centralities are plotted in Fig.~\ref{Fig5}. Centrality is represented by the fraction of total cross section $\sigma/\sigma_{\rm tot}$, where results for peripheral collisions are shown on the left-hand side. Lines connect the results for each model. The standard model function $\chi^2$ values are shown by star symbols, sextupole results by solid blue circles, same-side 1D azimuth Gaussian model by solid red squares, non-Gaussian exponent model [Eq.~(\ref{Eq13})] results by solid green triangles, $\eta_\Delta$-dependent polynomial [Eq.~(\ref{Eq14})] results by solid magenta diamonds, $\eta_\Delta$-dependent polynomial with non-Gaussian $\phi_\Delta$ exponent [Eq.~(\ref{Eq15})] by solid brown ``plus'' symbols, and the quartic-model [Eq.~(\ref{Eq16})] results by open cyan triangles. The left panel (a) of Fig.~\ref{Fig5} shows the absolute $\chi^2$/DoF values whose general trend with centrality is the same as that reported in Ref.~\cite{axialCI}. The differences in $\chi^2$/DoF relative to the standard model function results are shown in panel (b) for clarity.  

From these results we find that all of the non-Gaussian models reduce the $\chi^2$/DoF for the mid- to more-central collision data from $(1 - \sigma/\sigma_{\rm tot})$ = 0.4 to 0.9. The sextupole model is not special in that regard. For the NG models studied here the quartic model in Eq.~(\ref{Eq16}) produces the best overall $\chi^2$/DoF.  The $\eta_\Delta$-dependent polynomial models [Eqs.~(\ref{Eq14}) and (\ref{Eq15})] are competitive for mid- to most-central correlation data but are inferior for the more-peripheral data where the same-side peak is narrower on $\eta_\Delta$ and less well represented by a polynomial truncated at the $\eta_\Delta^4$ term. Note that the NG dependence of the same-side 2D peak model starts to improve the $\chi^2$/DoF near the sharp transition at mid-centrality ($\sigma/\sigma_{\rm tot} \approx 0.5$) but has diminished effect towards more-central bins where the $\eta_\Delta$ width exceeds the $\eta_\Delta$ acceptance (2 units)~\cite{STARTPC}. The latter may be a result of the non-Gaussian tails evident in panel (c) of Fig.~\ref{Fig6} being pushed outside the $\eta_\Delta$ acceptance.

Correlation measurements with higher $p_t$ cuts, or with so-called trigger-associated $p_t$ selection criteria~\cite{Joern,Dihadron1,Dihadron2,Dihadron3}, or for the higher collision energies attained at the LHC~\cite{atlas,cms,alice} provide strong evidence for non-Gaussian dependence in the same-side 2D correlation peak.  Therefore it should not be surprising if some small degree of non-Gaussian dependence exists for same-side $p_t$-integral correlations at RHIC energies. Such occurrence would not exclude the possibility that the same-side correlation peak is dominated by perturbative QCD jets with modified fragmentation~\cite{Tomjetfrag,Tommodfrag,BW}. It seems more plausible for possible non-Gaussian structure in these data to originate locally in relative azimuth rather than arising from the combination of a same-side peaked structure with global angle correlations, such as $m > 2$ harmonics. 

\section{Same-side model invariants}
\label{SecV}

In Refs.~\cite{axialCI,Tomv3-1,Tomv3-2} it was shown that the $\eta_\Delta$ projection of the same-side angular correlation data for minimum-bias 200 GeV Au+Au collision data can be well described by a single Gaussian or by a constant plus a reduced Gaussian. For the latter case the individual fitting parameters (constant, Gaussian amplitude and width) have large statistical uncertainties and large covariances compared to those of the single-Gaussian model. On the other hand, the large covariances enable the total amplitude (constant plus Gaussian) and curvature at the peak at $\eta_\Delta = 0$ to be determined accurately with small uncertainties. That example shows that the same basic properties of correlation structures in the data can be obtained from different fitting models by constructing {\em model invariants} based on the co-variation among the model parameters. In this section a similar analysis is carried out with the same-side 2D peak of the correlation data using two fitting models: a single 2D Gaussian and a sum of multipoles for $m \in [1,3]$ plus a reduced 2D Gaussian.

Examples are shown in Fig.~7 of Ref.~\cite{axialCI} for 38-46\% and 9-18\% 200 GeV Au+Au collision data, where the offset ($A_0$), dipole, quadrupole, and sharp 2D exponential ($A_{\rm bkg}$ term) of the standard model function in Eq.~(\ref{Eq4}) have been subtracted from the data. A similar subtraction was applied to the correlation data in each of the six more-central bins where the same-side correlation peak extends beyond the $\eta_\Delta$ acceptance edge. The subtracted model elements were considered to be fixed quantities in order that the resulting error matrices for parameters of the two fitting models correspond to uncertainties in the same-side peak shape only.

For the single 2D Gaussian fit the statistical uncertainties in the parameters $A_{\rm 2D}$, $\sigma_{\eta_\Delta}$, and $\sigma_{\phi_\Delta}$ and their covariances were much smaller than for the complete standard model function (see Table III in Ref.~\cite{axialCI}). The statistical uncertainties in the amplitudes varied between $\pm 0.0015$ and $\pm 0.0027$ for the six more-central bins compared to uncertainties in $A_{\rm 2D}$ for the standard model function which varied between $\pm 0.017$ and $\pm 0.050$.

For the fitted multipole + 2D Gaussian model the subtracted data were fitted with the following function
\bea
\label{Eq17}
F_{\rm sub}(\eta_\Delta,\phi_\Delta) & = & A_0^{\prime  \prime}
 + A_{\rm D}^{\prime  \prime} \cos(\phi_\Delta - \pi)/2 \nonumber \\
& & \hspace{-0.75in} +  2A_{\rm Q}^{\prime  \prime} \cos(2\phi_\Delta) 
   +   2A_{\rm S}^{\prime  \prime} \cos(3\phi_\Delta) \nonumber \\
& & \hspace{-0.75in}  +  A_{\rm 2D}^{\prime  \prime} \exp \left\{ - \frac{1}{2} \left[ \left( \frac{\eta_\Delta}{\sigma_{\eta_\Delta}^{\prime  \prime}} \right)^2 + \left( \frac{\phi_\Delta}{\sigma_{\phi_\Delta}^{\prime  \prime}} \right)^2 \right] \right\}.
\eea
The same-side effective ridge is determined by the first four terms with amplitude $B_{\rm amp}$ given by
\bea
\label{Eq18}
B_{\rm amp} & = & A_0^{\prime  \prime} - A_{\rm D}^{\prime  \prime}/2
 + 2A_{\rm Q}^{\prime  \prime} + 2A_{\rm S}^{\prime  \prime}
\eea
and with total variance
\bea
\label{Eq19}
\sigma^2_{B_{\rm amp}} & = & \sigma^2_{A_0^{\prime  \prime}}
  + \sigma^2_{A_{\rm D}^{\prime  \prime}}/4
  + 4\sigma^2_{A_{\rm Q}^{\prime  \prime}} + 4\sigma^2_{A_{\rm S}^{\prime  \prime}}
\eea
expressed as a sum of the variances of the amplitudes of the four contributing terms. The covariance of parameters $B_{\rm amp}$ and $A_{\rm 2D}^{\prime  \prime}$, the amplitude of the 2D Gaussian, is given by 
\bea
\label{Eq20}
cov(B_{\rm amp},A_{\rm 2D}^{\prime  \prime}) & = & 
cov(A_0^{\prime  \prime},A_{\rm 2D}^{\prime  \prime}) -
cov(A_{\rm D}^{\prime  \prime},A_{\rm 2D}^{\prime  \prime})/2 \nonumber \\
& & \hspace{-0.5in}  +  2cov(A_{\rm Q}^{\prime  \prime},A_{\rm 2D}^{\prime  \prime}) +
2cov(A_{\rm S}^{\prime  \prime},A_{\rm 2D}^{\prime  \prime}).
\eea
The amplitude of the same-side effective ridge is determined by the sextupole in conjunction with the away-side data constraint.

\begin{figure*}[t]
\includegraphics[keepaspectratio,width=2.25in]{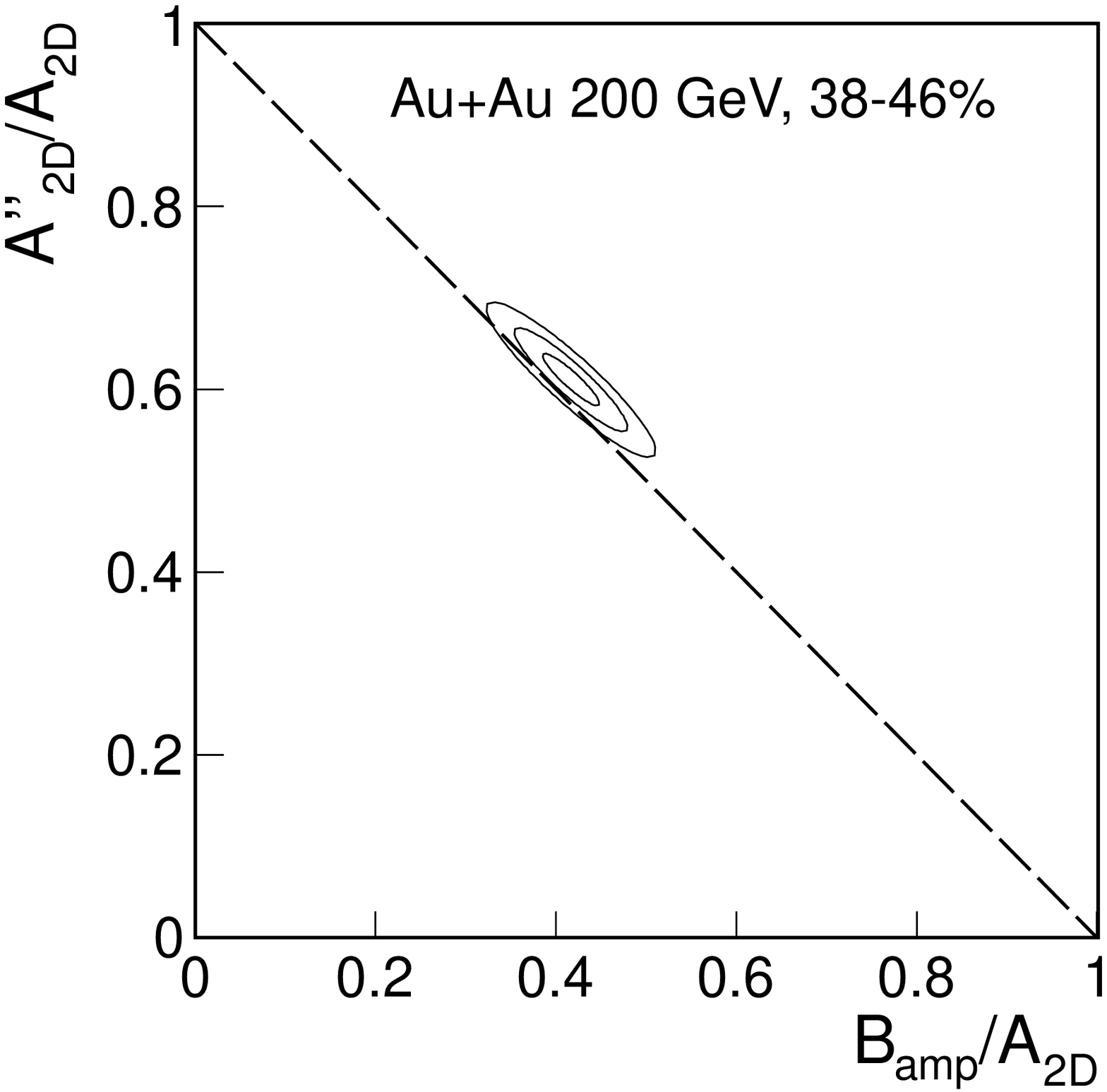}
\put(-125,30){\bf (a)}
%\put(-90,25){\bf 38-46\%}
\includegraphics[keepaspectratio,width=2.25in]{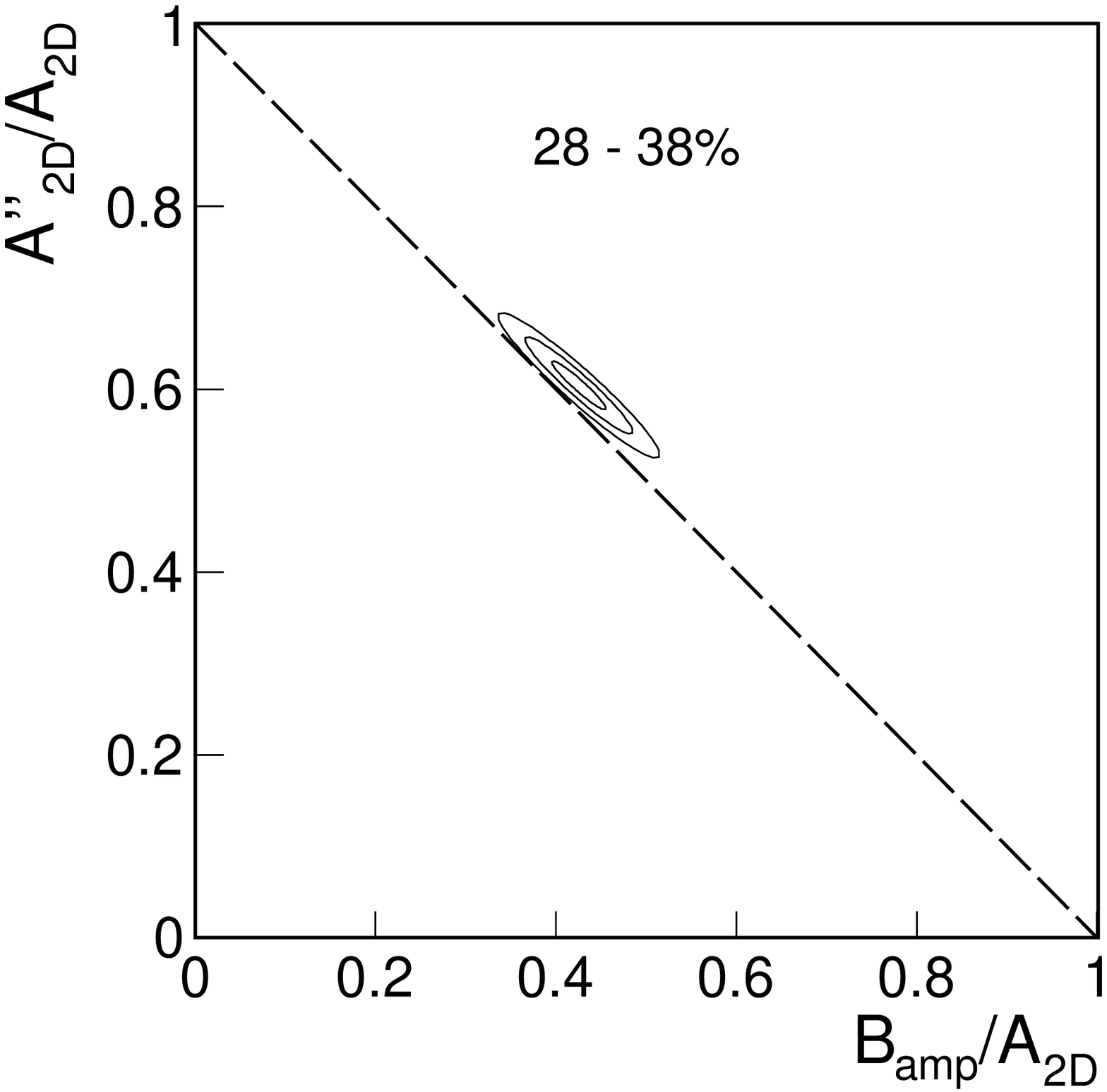}
\put(-125,30){\bf (b)}
%\put(-90,25){\bf 28-38\%}
\includegraphics[keepaspectratio,width=2.25in]{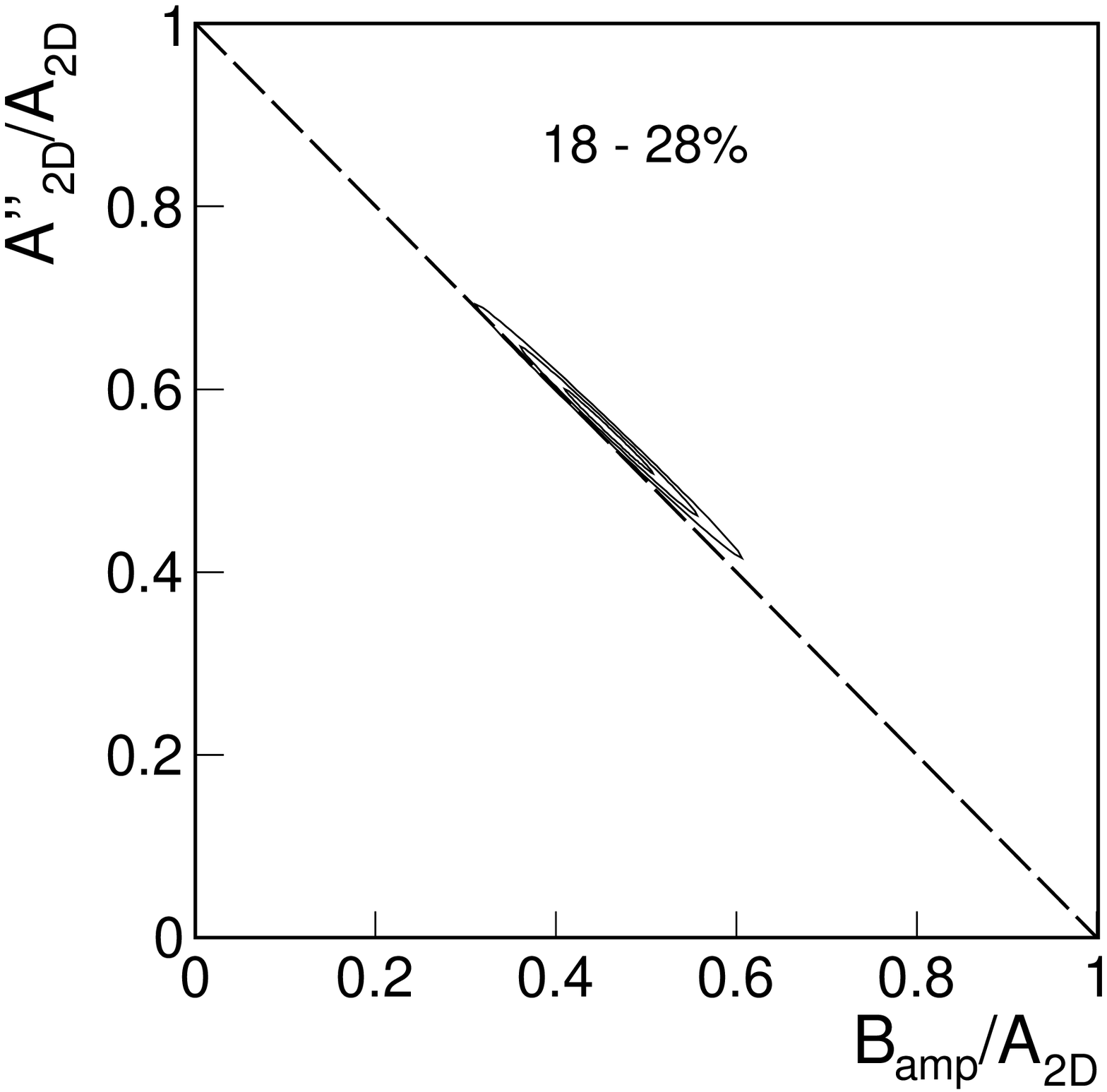}
\put(-125,30){\bf (c)}
%\put(-90,25){\bf 18-28\%}
\linebreak
\includegraphics[keepaspectratio,width=2.25in]{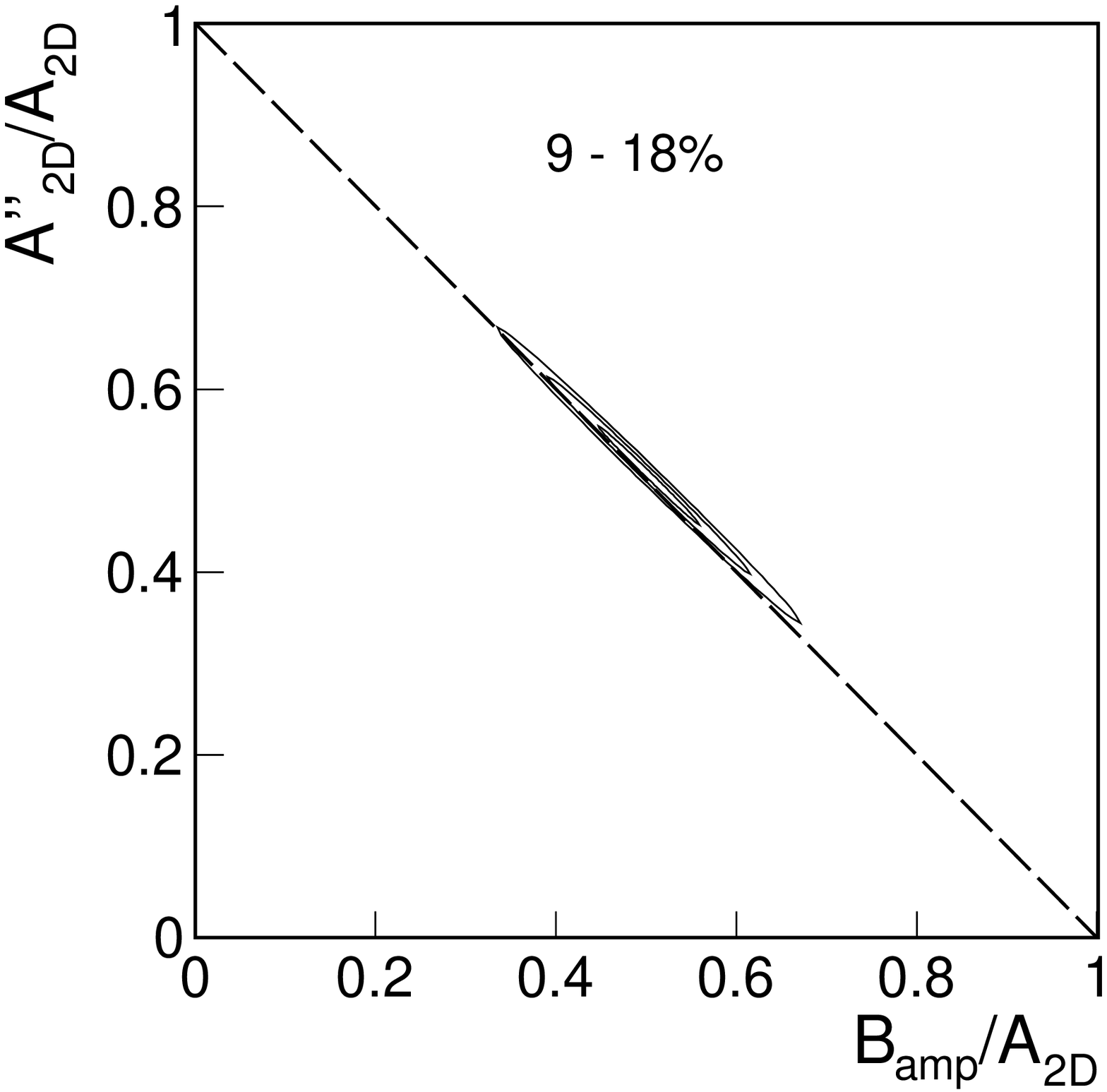}
\put(-125,30){\bf (d)}
%\put(-90,25){\bf 9-18\%}
\includegraphics[keepaspectratio,width=2.25in]{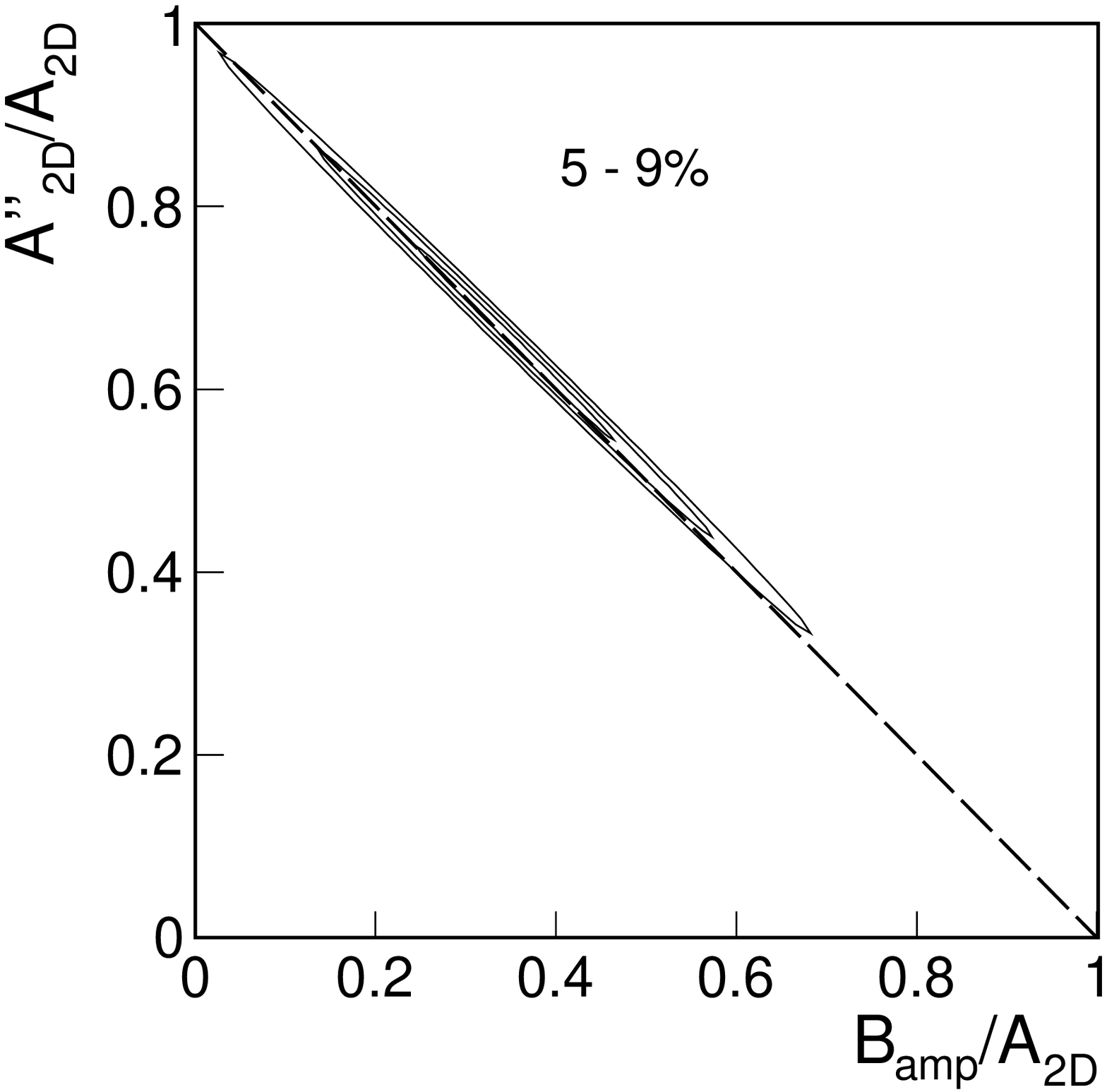}
\put(-125,30){\bf (e)}
%\put(-90,25){\bf 5-9\%}
\includegraphics[keepaspectratio,width=2.25in]{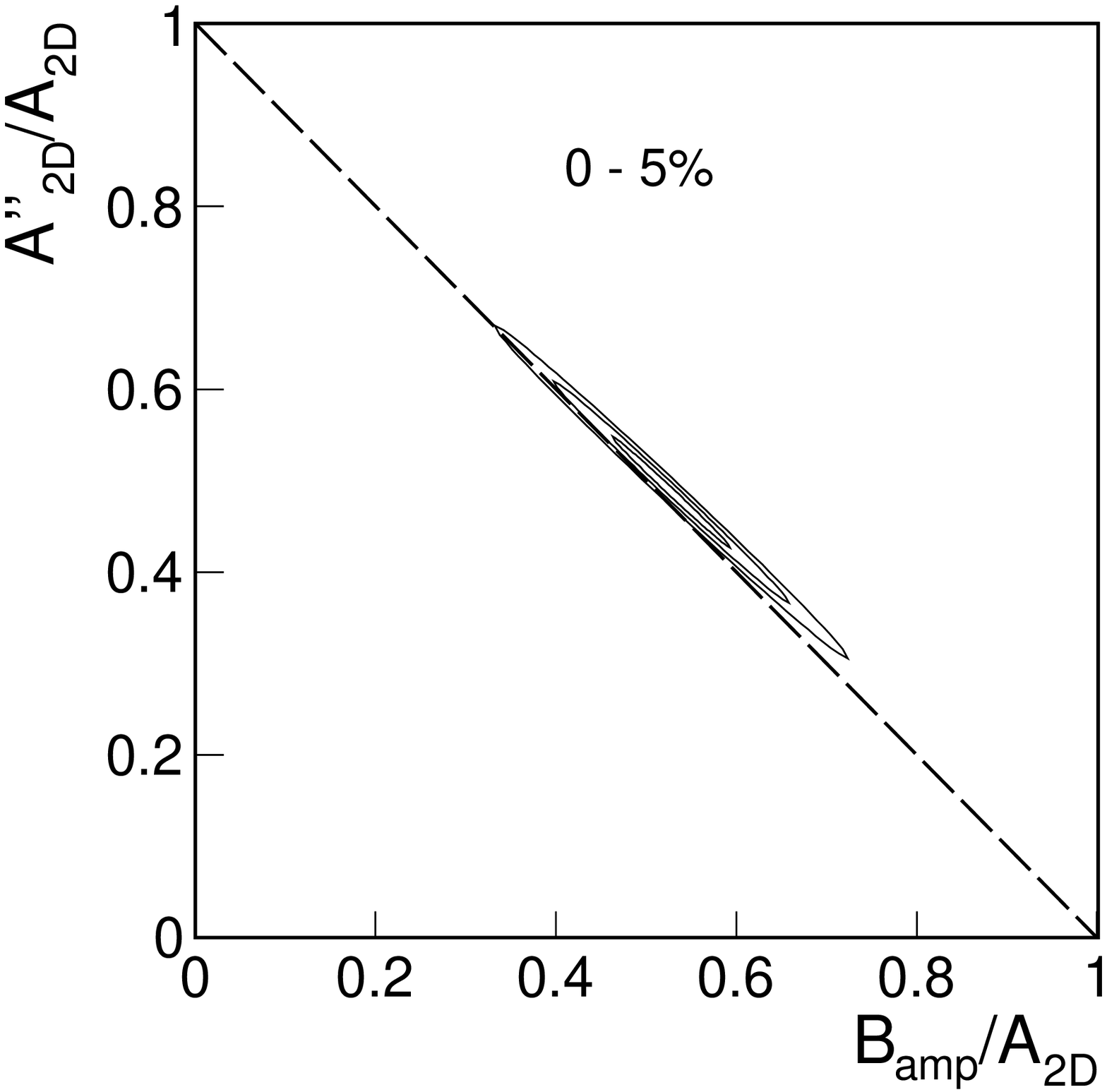}
\put(-125,30){\bf (f)}
%\put(-90,25){\bf 0-5\%}
\caption{\label{Fig7}
Parameter error contours (1, 2, and 3$\sigma$) for the model function in Eq.~(\ref{Eq17}) applied to the same-side subtracted angular correlation data for 200 GeV Au+Au collisions~\cite{axialCI} discussed in the text. Uncertainties are shown for the 2D Gaussian amplitude $A_{\rm 2D}^{\prime  \prime}$ versus those for the same-side effective ridge amplitude $B_{\rm amp}$ in Eq.~(\ref{Eq18}) due to the sextupole element. Both amplitudes are plotted as ratios to the single-Gaussian amplitude $A_{\rm 2D}$. The dashed lines represent the condition that $A_{\rm 2D}^{\prime  \prime} + B_{\rm amp} = A_{\rm 2D}$. Results are shown in panels (a) - (f) for centralities 38-46\%, 28-38\%, 18-28\%, 9-18\%, 5-9\% and 0-5\% respectively.}
\end{figure*}

The resulting amplitudes and errors are listed in Table~\ref{TableI}, and the 1, 2, and 3$\sigma$ error ellipses are shown in Fig.~\ref{Fig7} for the amplitude of the same-side 2D reduced Gaussian ($A_{\rm 2D}^{\prime  \prime}$) versus the amplitude of the same-side effective ridge ($B_{\rm amp}$). Both amplitudes are plotted relative to the single 2D Gaussian amplitude $A_{\rm 2D}$ for each centrality. Strong anticorrelations are evident in the error contours. The results in Table~\ref{TableI} and Fig.~\ref{Fig7} show that
\bea
\label{Eq21}
B_{\rm amp} + A_{\rm 2D}^{\prime  \prime} & \approx & A_{\rm 2D}
\eea
as expected, where for the more-central data the error contours accurately coincide with the reference sum $B_{\rm amp}/A_{\rm 2D} + A_{\rm 2D}^{\prime  \prime}/A_{\rm 2D} = 1$ (dashed lines in Fig.~\ref{Fig7}). The statistical uncertainties in the amplitudes of the same-side azimuth ridge and the reduced 2D Gaussian are roughly an order of magnitude larger than the corresponding uncertainties in the single 2D Gaussian amplitudes. The statistical uncertainties along the sum directions in Fig.~\ref{Fig7} (semi-minor axis length of each $1\sigma$ error ellipse) listed in the last column of Table~\ref{TableI} are approximately the same as the uncertainties in the single 2D Gaussian amplitude.

The strong anticorrelations illustrate the inherent ambiguity in fits to the same-side 2D peak structure when the model permits a uniform ridge on $\eta_\Delta$ plus a reduced amplitude 2D Gaussian. Those anticorrelations increase for more-central collisions where the $\eta_\Delta$ width of the same-side 2D peak is largest. The combined results in Figs.~\ref{Fig5} and \ref{Fig7} show that for the more-central collision data $\chi^2$ dependence on the sextupole amplitude or its proxy $B_{\rm amp}$ is weak relative to mid-centrality. This reduced $\chi^2$ sensitivity to NG dependence occurs when $\sigma_{\eta_\Delta} > \Delta\eta$ which reduces the relative magnitude of the $\eta_\Delta^4$ term in Eq.~(\ref{Eq12}); the $\eta_\Delta^4$ term contains the leading-order NG contribution. The results in this section demonstrate the inherent fitting instability resulting from introduction of superfluous model elements, in this case the sextupole.

%%%%%%%%%%%%%%%%%%%%%%%%%%%%%%%%%%
\begin{table*}[t]
\caption{Comparison of results for model descriptions of the subtracted angular correlation data for 200 GeV Au+Au minimum-bias collisions~\cite{axialCI}. The first column lists centrality as percent of total cross section; the remaining columns from left-to-right list the results of fits for the single-Gaussian amplitude, the same-side effective ridge, the reduced 2D Gaussian amplitude, the uncertainties for the latter two amplitudes, and the amplitude sum $(B_{\rm amp} + A_{\rm 2D}^{\prime  \prime})$. Statistical fitting uncertainties are in parentheses.}
\label{TableI}
\begin{tabular}{cccccc}
\hline \hline
Centrality & $A_{\rm 2D}$ & $B_{\rm amp}$ & $A_{\rm 2D}^{\prime  \prime}$ & $\pm$uncertainty  &  $(B_{\rm amp} + A_{\rm 2D}^{\prime  \prime})$ \\
   (\%) & & & & $\times 10^4$ \\
\hline
38-46  &  0.3179(15)  &  0.1325  &  0.1941  &  93  &  0.3266(27) \\
28-38  &  0.3725(16)  &  0.1585  &  0.2251  & 104  &  0.3836(25) \\
18-28  &  0.5820(15)  &  0.2669  &  0.3226  & 278  &  0.5895(17) \\
 9-18  &  0.7570(16)  &  0.3808  &  0.3830  & 416  &  0.7638(26) \\
 5-9   &  0.7417(24)  &  0.2630  &  0.4824  & 798  &  0.7454(35) \\
 0-5   &  0.6432(27)  &  0.3395  &  0.3136  & 405  &  0.6531(25) \\
\hline \hline
\end{tabular}
\end{table*}
%%%%%%%%%%%%%%%%%%%%%%%%%%%%%%%

%%%%%%%%%%%%%%%%%%%%%%%%%%%%%%%%%%
%%%%%%%%%%%%%%%%%%%%%%%%%%%%%%%%%%

\section{ATLAS Pb+Pb 0-1\% correlation data}
\label{SecVI}

Minimum-bias Pb+Pb collision data at $\sqrt{s_{\rm NN}}$ = 2.76 TeV with corresponding 2D angular correlations have recently been reported by three major experiments~\cite{atlas,cms,alice} at the Large Hadron Collider (LHC).  The 2D correlations are quite interesting, not only because of the unprecedented high collision energy and wide pseudorapidity acceptance (CMS and ATLAS) but also because the most-central correlation data reveal an away-side double-peaked structure on azimuth~\cite{atlas,alice}. Na\"{i}ve interpretation of this double-peaked structure might include a same-side azimuth sextupole. In this section the fitting models introduced in Secs.~\ref{SecII} and \ref{SecIV} are applied to the 2D angular correlations for the 0-1\% centrality Pb+Pb collision data from the ATLAS experiment~\cite{atlas} where the data include all charged particles with $2 \leq p_t \leq 3$~GeV/$c$.

%%%%%%%%%%%%%%%%%%%%%%%%%%%%%%%%%
\begin{figure*}[t]
\includegraphics[keepaspectratio,width=1.8in]{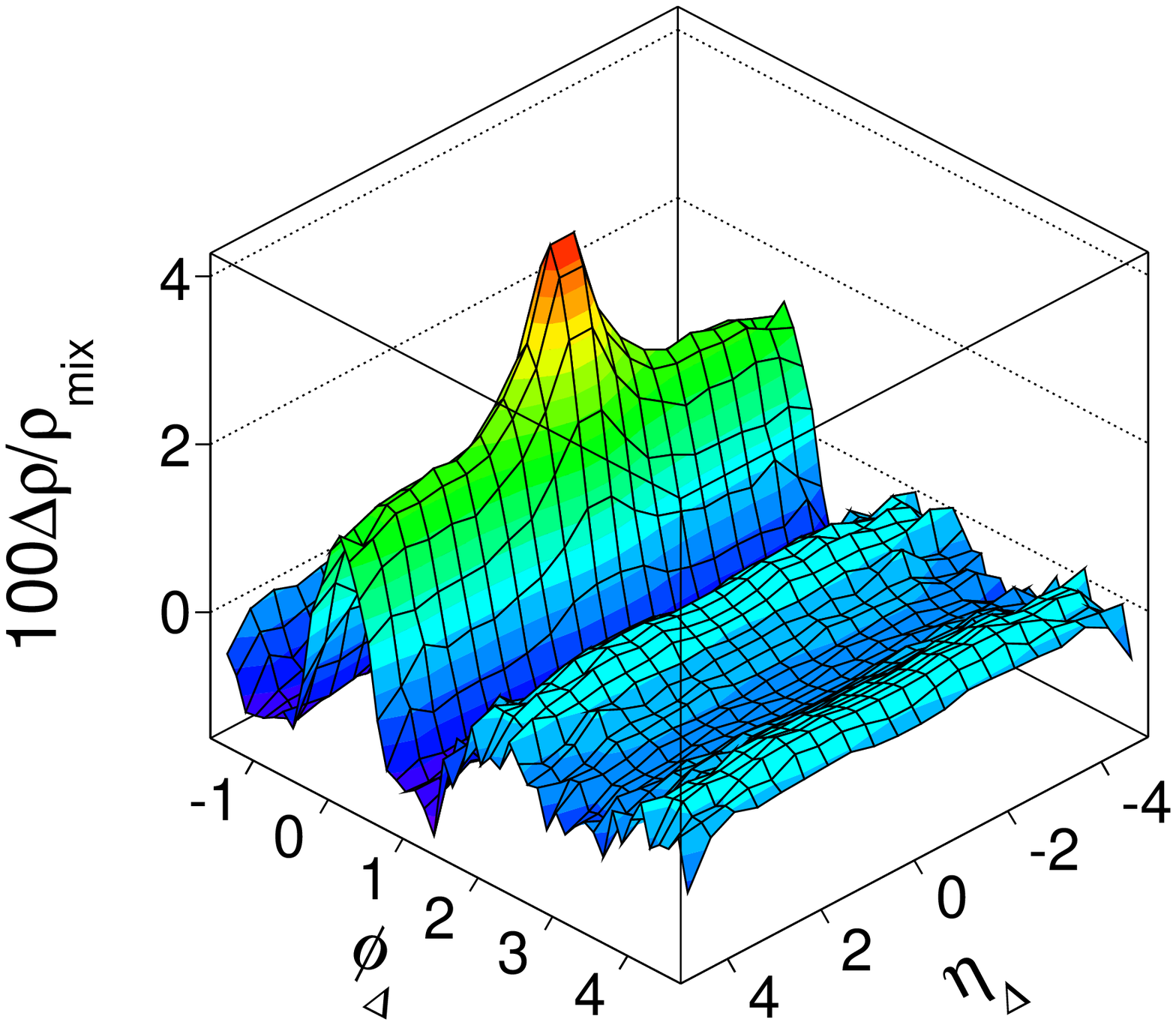}
\put(-75,93){\bf (a)}
\includegraphics[keepaspectratio,width=1.8in]{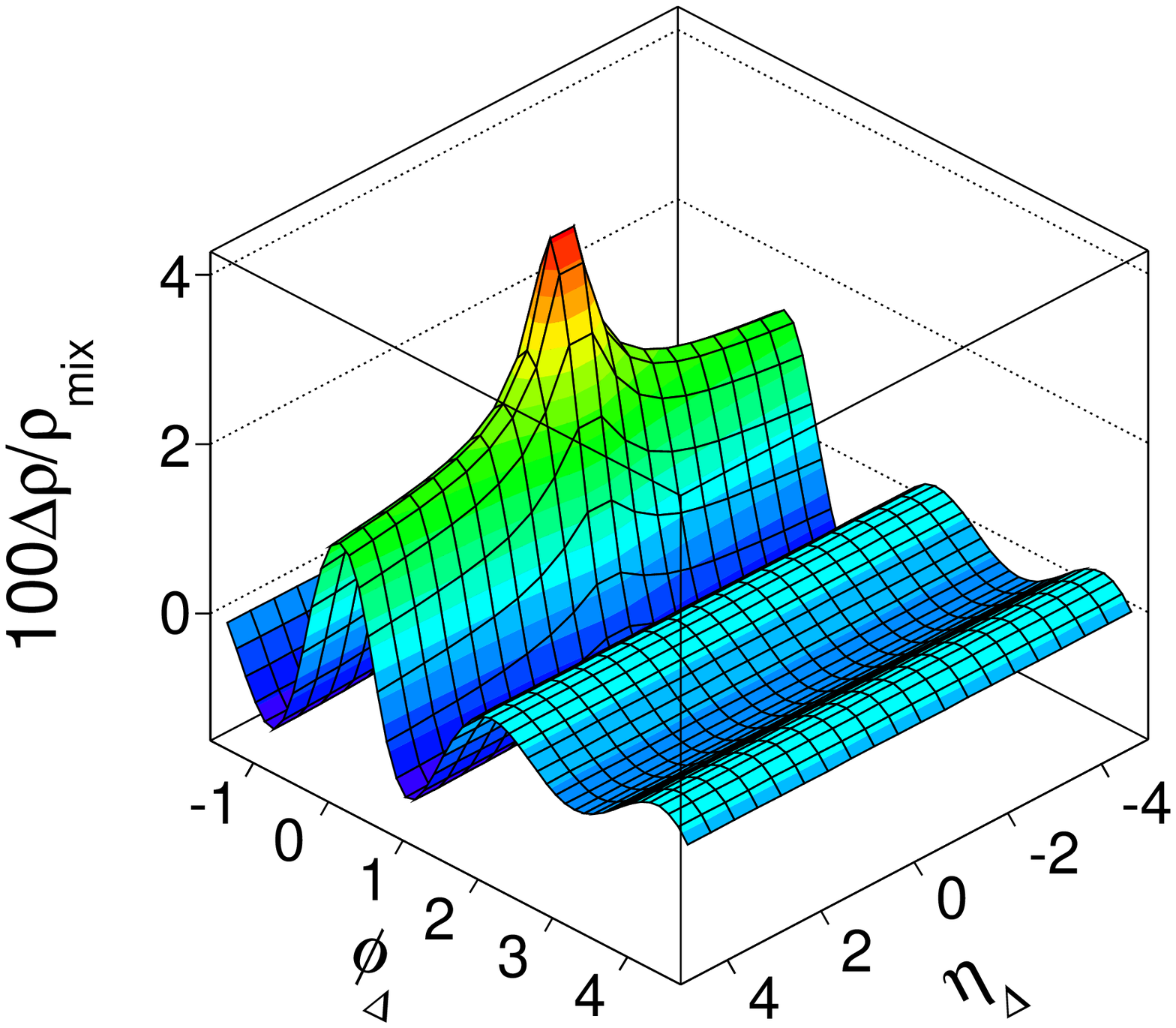}
\put(-75,93){\bf (b)}
\includegraphics[keepaspectratio,width=1.8in]{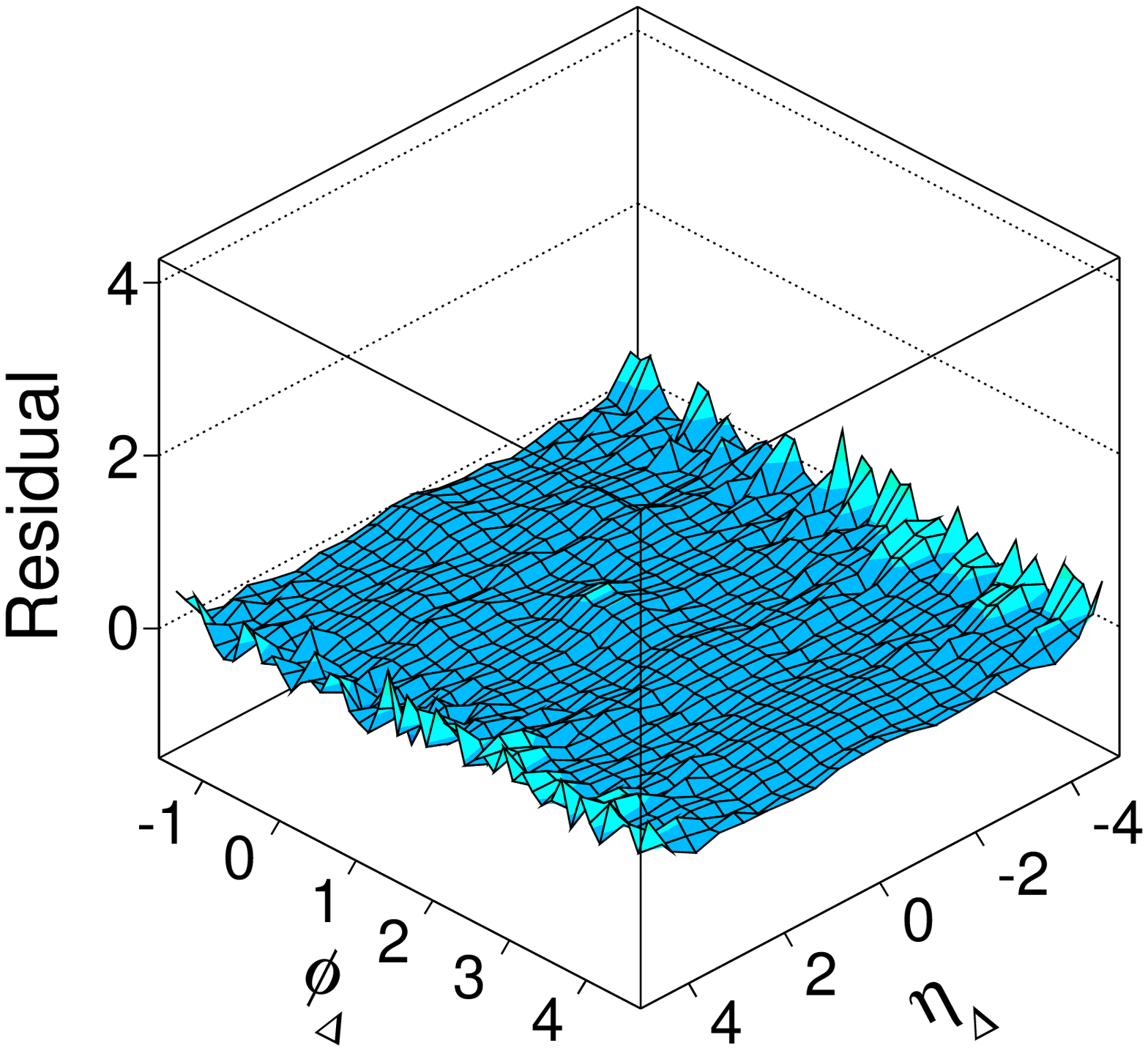}
\put(-75,93){\bf (c)}
\includegraphics[keepaspectratio,width=1.8in]{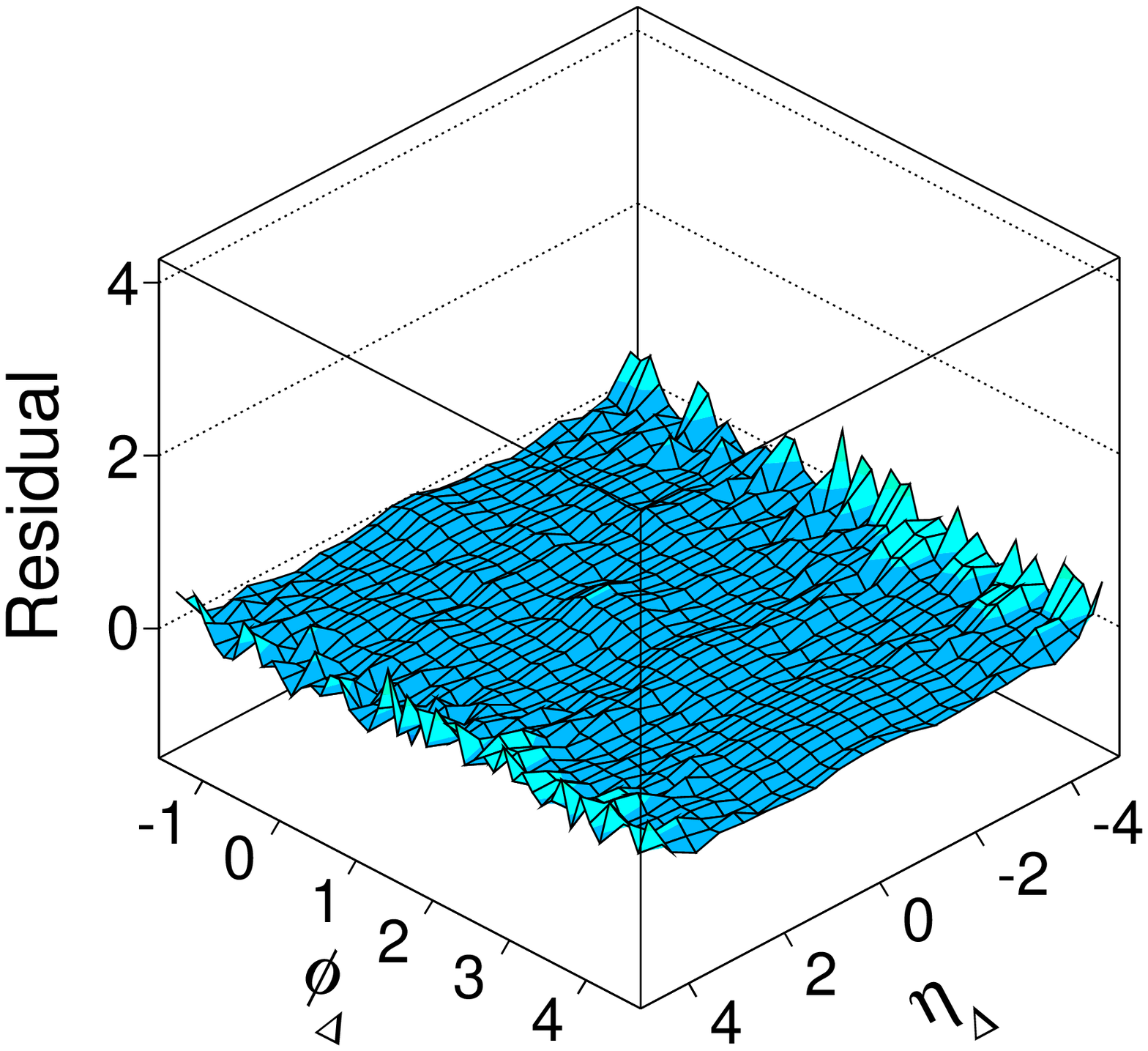}
\put(-75,93){\bf (d)}
\linebreak
\includegraphics[keepaspectratio,width=1.8in]{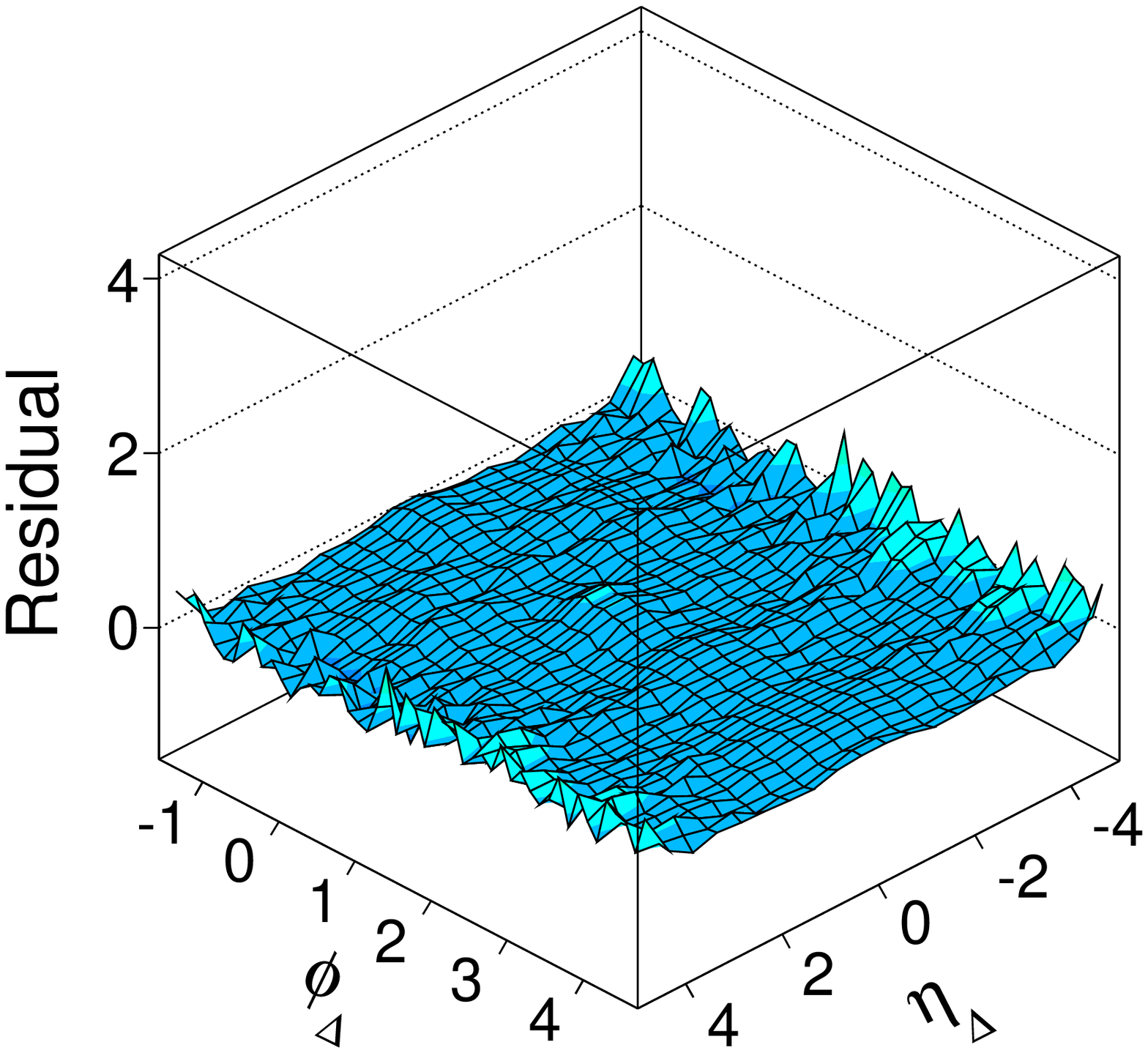}
\put(-75,93){\bf (e)}
\includegraphics[keepaspectratio,width=1.8in]{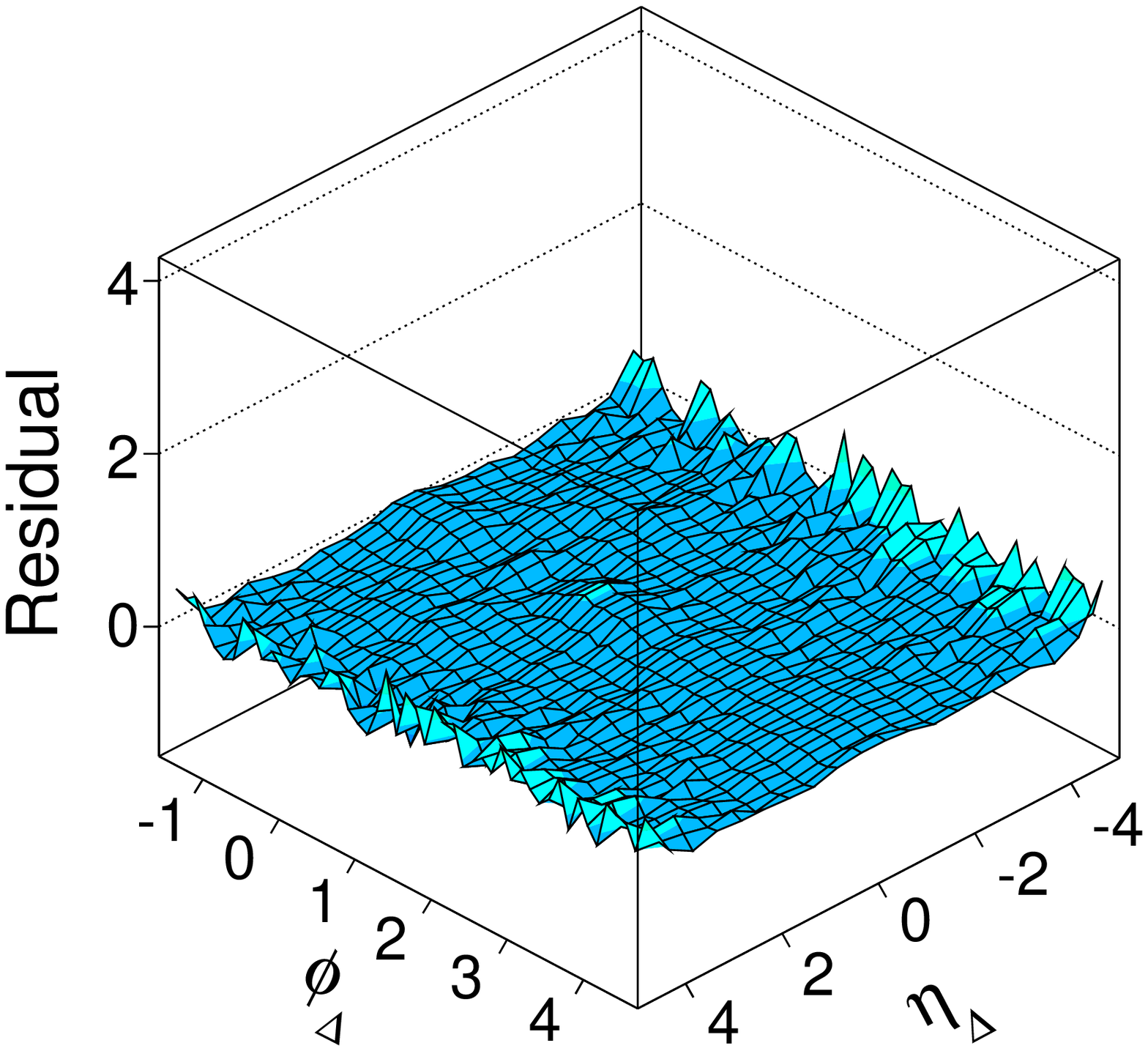}
\put(-75,93){\bf (f)}
\includegraphics[keepaspectratio,width=1.8in]{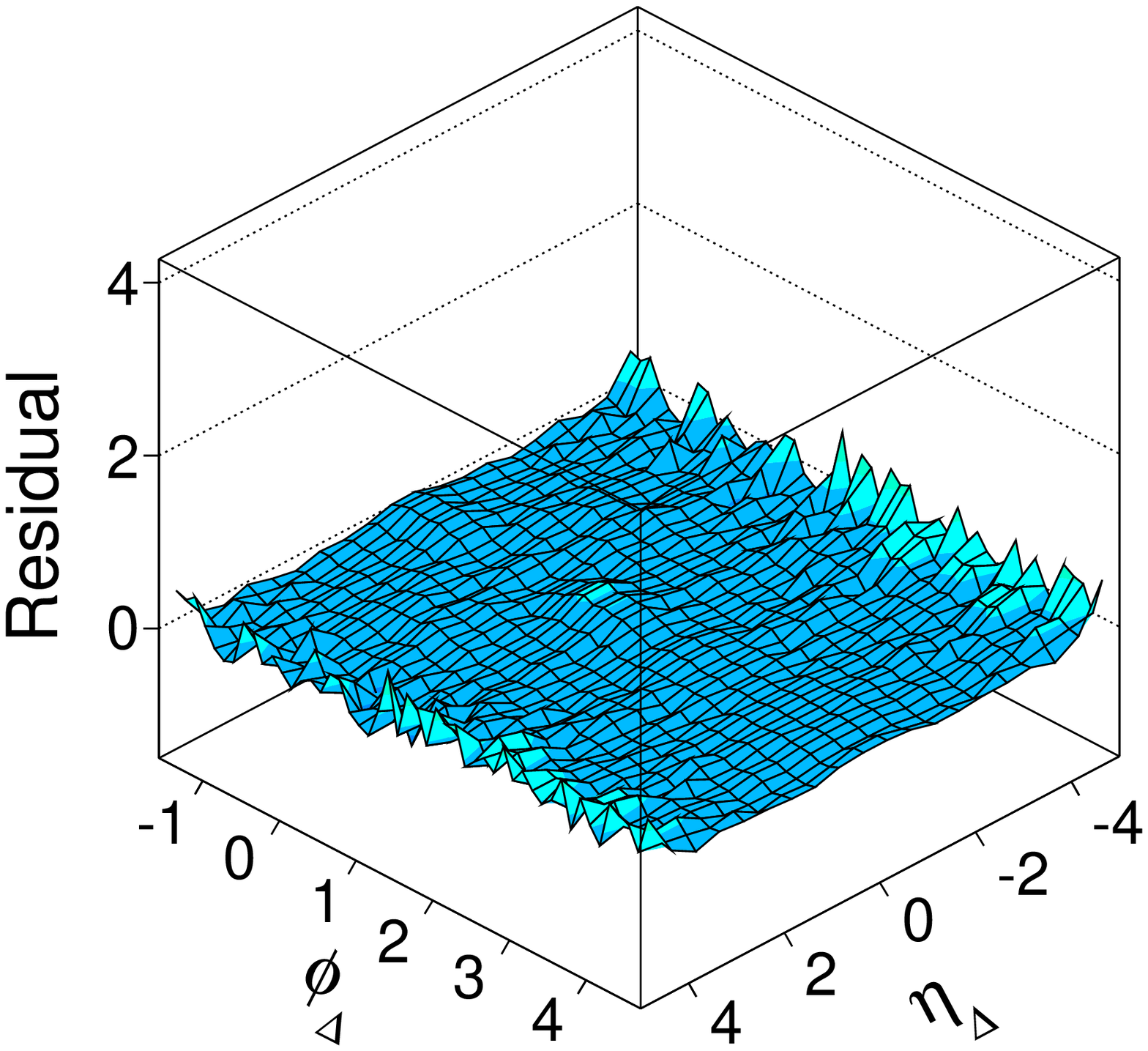}
\put(-75,93){\bf (g)}
\caption{\label{Fig8}
(Color online) Panel (a): angular correlation data for 0-1\% centrality 2.76 TeV Pb+Pb collisions from ATLAS~\cite{atlas}; panel (b): typical best fit assuming the Quad(a) set of model elements discussed in the text; panels (c) - (g): residuals (model - data) for the fitting models denoted by Quad(a), Quad(b), ASG, Q\&S, and ASG\&S respectively as discussed in the text.}
\end{figure*}
%%%%%%%%%%%%%%%%%%%%%%%%%%%%%%%%%%

The same-side correlations decrease monotonically with $\eta_\Delta$ from the peak at $\eta_\Delta = 0$ to the $\eta_\Delta$ = 5 acceptance limit while maintaining a significant positive value.  This structure is most likely non-Gaussian and can be described as a single peaked function, as a sum of two peaked functions or as the sum of a peaked function plus a uniform same-side ridge. 
The away-side double-peaked structure in the data may require a negative quadrupole, a positive same-side sextupole, a negative away-side 1D Gaussian or a combination of these model elements. The phenomenological fitting model we use to describe these data includes the elements in Eq.~(\ref{Eq4}) plus a sextupole with variable exponents for the same-side 2D peak function as in Eq.~(\ref{Eq13}) plus a periodic away-side 1D Gaussian (ASG) on azimuth given by
\bea
\label{Eq22}
F_{\rm ASG}(\eta_\Delta,\phi_\Delta) & = & A_{\rm ASG} \hspace{-0.1in} \sum_{k=-3,-1\cdots5} \hspace{-0.1in}
e^{-(\phi_\Delta - k\pi)^2/2\sigma^2_{\rm AS}}.
\eea
The ASG model element accounts for the possibility that the dip in the data on $\phi_\Delta$ at $\phi_\Delta$ = $\pi$ could be due to physical processes which act locally on azimuth, for example attenuation of back-to-back correlated pairs from dijets.  

The ATLAS correlation data were reported as ratio $\rho_{\rm sib}/\rho_{\rm mix}$. Conversion to the per-particle normalization described in Sec.~\ref{SecII} requires $dN_{\rm ch}/d\eta$ for the $p_t$ interval [2,3]~GeV/$c$ which are not available~\cite{ATLASspec}. The ATLAS correlation data were arbitrarily normalized as
\bea
\label{Eq23}
100 \left( \frac{\rho_{\rm sib}}{\rho_{\rm mix}}|_{\rm ATLAS} - 1 \right)
& = & 100 \Delta\rho/\rho_{\rm mix}|_{\rm ATLAS}
\eea
for convenience. The normalization factor affects the amplitudes only, not the shape of the angular correlations which is of primary interest. The resulting correlations are shown in panel (a) of Fig.~\ref{Fig8}.

Descriptions of the data were attempted using a variety of fitting-model options.  In all cases the away-side dipole, the same-side 2D peak function ($A_{\rm 2D}$ element), and the 2D exponential ($A_{\rm bkg}$ element) in Eq.~(\ref{Eq4}) were included. The exponents for the $A_{\rm 2D}$ term
  were allowed to vary in each case, as defined in Eq.~(\ref{Eq13}). The soft-component model element ($A_{\rm soft}$ term) was not required in this case because of the $p_t$ cuts, consistent with the earlier discussion about the $y_{t1} + y_{t2}$ cuts. The fitting options were the following: (i) Include each of the quadrupole, sextupole and away-side 1D Gaussian separately. (ii) Combine the quadrupole and sextupole simultaneously in the fit. (iii) Combine the sextupole and ASG simultaneously. Fitting models in which the quadrupole and ASG were included or in which all three model elements were simultaneously included were unstable and displayed strong covariation between amplitudes.

Results assuming a quadrupole only (no sextupole or ASG) yielded negative quadrupole amplitudes as expected, and as large as $-0.48$ (refer to the correlation structure in Fig.~\ref{Fig8}a). Distinct results were obtained depending on the choice of shape for the same-side 2D exponential element, where the exponent was either fixed to 1/2 or allowed to vary. For the latter option that term in Eq.~(\ref{Eq4}) was generalized to
\bea
\label{Eq24}
A_{\rm sharp} \exp \left\{ - \left[ \left( \frac{\eta_\Delta}{w_{\eta_\Delta}} \right)^2 
+ \left( \frac{\phi_\Delta}{w_{\phi_\Delta}} \right)^2 \right]^{\xi/2} \right\}
\eea
with variable exponent $\xi/2$. For fixed exponent ($\xi = 1$) both the $A_{\rm 2D}$ and $A_{\rm sharp}$ model elements produced peaks which combined to describe the same-side data. When the exponent $\xi$ was allowed to vary the 2D exponential function described almost all of the 2D peak shape in the data while the $A_{\rm 2D}$ term described an approximately uniform same-side ridge. The respective $\chi^2$/DoF values were 3.05 and 2.81 and the respective model parameter values are listed in Table~\ref{TableII} with the labels ``Quad(a)'' and ``Quad(b).'' The Quad(a) fitted model function is shown in panel (b) of Fig.~\ref{Fig8}; residuals for the Quad(a) and Quad(b) fits are shown in Fig.~\ref{Fig8} panels (c) and (d) respectively. Fits in which the $A_{\rm sharp}$ model element was omitted were poor.

The sextupole only (no quadrupole or ASG) fitting results obtained a positive same-side sextupole amplitude as expected from the shape of the away-side data, where $A_{\rm S}$ = 0.17.  However the fit quality was poor, with large $\chi^2$/DoF = 7.3 compared with other fits discussed in this section. For this fitting solution both of the same-side 2D peak model elements had large magnitudes (larger than the structures in the data) with opposite algebraic signs. The same-side correlation peak was therefore described with large, canceling model elements, a non-intuitive fitting solution. 

The model description using the away-side Gaussian only (no quadrupole or sextupole) provided good fits to the data where $\chi^2$/DoF = 3.24. However, the away-side dipole and Gaussian model elements strongly co-varied, resulting in large canceling amplitudes relative to that of the data with values $A_{\rm D}$ = 13.49 and $A_{\rm ASG}$ = $-10.89$. The residuals for this case are shown in Fig.~\ref{Fig8}e and the parameter values are listed in Table~\ref{TableII} with the label ``ASG.''

The data were also successfully described when the fitting model included all three azimuth multipoles $m$ = 1, 2 and 3. However, the optimum fits and even the fitting stability depended on assumptions about the $A_{\rm sharp}$ model element. If the function in Eq.~(\ref{Eq24}) was restricted by requiring $\xi$ = 1, then both it and the $A_{\rm 2D}$ term contributed to the shape of the same-side peak resulting in the best fit for the three multipole fitting model where $\chi^2$/DoF = 2.90. The obtained parameter values were $A_{\rm D}$ = 4.90, $A_{\rm Q}$ = $-0.69$ and $A_{\rm S}$ = $-0.078$. Note that both the quadrupole and sextupole are negative. The remaining parameter values are listed in Table~\ref{TableII} under heading ``Q\&S.'' The residuals are shown in Fig.~\ref{Fig8}f.

%%%%%%%%%%%%%%%%%%%%%%%%%%%%%%%%%%
\begin{table}[htb]
\caption{Model function parameters for the ATLAS Pb+Pb 0-1\% centrality angular correlation data~\cite{atlas}. The model functions and results discussed in the text using the parameters defined in Eqs.~(\ref{Eq4}), (\ref{Eq13}), (\ref{Eq22}) and (\ref{Eq24}) are listed in the columns under the model labels introduced in the text.}
\label{TableII}
\begin{tabular}{lccccc}
\hline \hline
Parameter & Quad(a) & Quad(b) & ASG & Q\&S & ASG\&S    \\
\hline
$A_{\rm D}$          &  3.40  &  3.46  &  13.49  &  4.90  &  13.54  \\
$A_{\rm Q}$           & $-$0.47 & $-$0.48 & 0.0  & $-$0.69 &   0.0   \\
$A_{\rm S}$           &  0.0   &  0.0   &   0.0  & $-$0.078 &  $-$0.157  \\
$A_{\rm 2D}$           & 8.44  & 5.49  &  6.76 & 12.26  &  10.75  \\
$\sigma_{\eta_\Delta}$ &  4.68  &  9.73  &  13.71 &  0.62  &   7.73  \\
$\sigma_{\phi_\Delta}$ &  0.480 &  0.483 &  0.471 &  0.512 &   0.504 \\
$\gamma$               & 0.0657 &  1.379 &  0.1318 & 0.0293 &  0.0453 \\
$\delta$               &  1.092 &  1.075 &  1.109  &  1.050 &   1.055 \\
$A_{\rm sharp}$        &  2.27  &  2.21  &  2.23   &  2.50  &   2.26  \\
$w_{\eta_\Delta}$      &  0.518 &  0.792 &  0.463  &  0.547 &   0.562 \\
$w_{\phi_\Delta}$      &  0.259 &  0.449 &  0.242  &  0.278 &   0.280 \\
$\xi$                  &  1.0   &  1.608 &  1.0    &  1.0   &   1.0   \\
$A_{\rm ASG}$          &  0.0   &  0.0   & $-$10.89 &  0.0   & $-$9.29  \\
$\sigma_{\rm AS}$      &  $-$   &  $-$   &  1.104  &  $-$   &   0.903 \\
$A_{\rm 0}$            & $-$2.69  & $-$2.74  & $-$8.48 & $-$3.89 & $-$9.45 \\
$\chi^2/DoF$           &  3.05  &  2.81  &  3.24   &  2.90  &   2.99  \\
\hline \hline
\end{tabular}
\end{table}
%%%  The runs are 7NG1, 7NG6, 7NG12, 7NG4, 7NG11  %%%
%%%%%%%%%%%%%%%%%%%%%%%%%%%%%%%

Other solutions were possible for the three-multipole fitting model discussed in the preceding paragraph. Allowing $\xi$ in Eq.~(\ref{Eq24}) to vary resulted in that model element conforming entirely to the shape of the same-side peak in the data while the $A_{\rm 2D}$ term was an approximately uniform ridge. In this situation the overall fitting model had, in effect, four independent functions with which to describe the $\eta_\Delta$-independent features of the data, an under-constrained model fit. This condition destabilized the fits causing strong parameter co-variations and continuous parameter ambiguities. Lastly, setting $A_{\rm sharp}$ to zero forced the $A_{\rm 2D}$ term with exponent variation [parameters $\gamma$ and $\delta$ in Eq.~(\ref{Eq13})] to conform to the peak structure in the data. For this solution $A_{\rm S}$ = 0.016, a positive sextupole amplitude. However, the $\chi^2$/DoF (= 4.85) was rather poor compared to typical values ($\approx$3) resulting from other model choices.

The model phenomenology including a dipole, sextupole and ASG (no quadrupole) was also successful. Inspection of the $\chi^2$ space for this choice of model elements found several local minima, each having competitive $\chi^2$ values.  These local minima appeared most clearly with respect to the variable exponent $\gamma$ in Eq.~(\ref{Eq13}) for the $\eta_\Delta$-dependent factor of the $A_{\rm 2D}$ term. The discrete fitting solutions describe qualitatively different shapes for the $\eta_\Delta$-dependence of the $A_{\rm 2D}$ term. These shapes correspond to a peaked distribution or one that is approximately constant. The latter resulted in an under-constrained fitting condition (four $\eta_\Delta$-independent functions), large parameter co-variation and instability.  The discrete solution with the smaller value of $\gamma$ = 0.045 produced a peaked 2D distribution, stable results, competitive $\chi^2$/DoF = 2.99, and the small residuals shown in Fig.~\ref{Fig8}g. The amplitude values were $A_{\rm D}$ = 13.54, $A_{\rm S}$ = $-0.157$, and $A_{\rm ASG}$ = $-9.29$. The dipole and ASG amplitudes are large and tend to cancel. The best-fit parameter values are given in Table~\ref{TableII} under heading ``ASG\&S.''

These results demonstrate that the 0-1\% Pb+Pb angular correlation data from ATLAS can be described phenomenologically with a variety of model functions. Those including a quadrupole obtain negative amplitudes which account for the dip in the away-side structure on $\phi_\Delta$. It is surprising that successful models which include a sextupole obtain a {\em negative} amplitude given the away-side double-peaked structure in the data which seems to suggest a positive same-side sextupole contribution. Models which include an away-side Gaussian obtain negative amplitudes as expected. However, the resulting away-side dipole and Gaussian amplitudes become quite large in magnitude relative to the structures in the data, producing strong cancellations. The $\chi^2$/DoF and residuals are comparable for each case; no one fitting model can be singled out as clearly preferred by the data.  Although visual inspection of the 0-1\% ATLAS data suggests the presence of a positive sextupole contribution, the present analysis excludes this conclusion. In general, the data do not require the presence of multipoles greater than $m$ = 1.

\section{Summary and Conclusions}
\label{SecVII}

Higher azimuth harmonics or $\cos(m\phi_\Delta)$ model elements have been introduced in the phenomenological description of angular correlation data from relativistic heavy-ion collisions to justify a hydrodynamic explanation for the $\eta_\Delta$-extended same-side 2D peak. The correlation data were described with fitted model parameters including an azimuth sextupole and even higher-order ($m > 3$) harmonics. However, advocates of higher harmonics have not demonstrated the necessity of these additional model elements and have not addressed the resulting instabilities in the fitted parameters.

In previous studies it was shown that a sextupole element was not required to describe the 2D angular correlations for 62 and 200 GeV Au+Au minimum-bias collisions from STAR~\cite{axialCI,Tomv3-1,Tomv3-2}. It was also shown that the inferred sextupole amplitudes followed directly from a Fourier series decomposition of the azimuth projection of the same-side 2D peak. These studies also showed that the net effect of including a sextupole element in the fit model was to replace part of the same-side 2D peak in the data with a same-side 1D effective ridge on azimuth. This effective-ridge contribution, when combined with the reduced 2D same-side peak function, was shown to be statistically equivalent to a single 2D Gaussian model such as that used in a previous analysis of STAR data~\cite{aya,axialCI}. The minor differences between the two descriptions of the same-side peak, though not systematically significant, lead to reduced $\chi^2$ values. In the present study those minor differences, the only surviving effect of the included sextupole, are shown to correspond to small non-Gaussian dependence in the same-side 2D peak. 

In the present study we show that one-dimensional projections of the correlation data onto $\eta_\Delta$ can be described with a model-independent polynomial function. The projections are consistent with a single-Gaussian hypothesis, but small non-Gaussian dependence can be introduced which further reduces $\chi^2$. We apply this idea to 2D angular correlation data and show that several choices for non-Gaussian same-side 2D peak functions, including a sextupole with reduced 2D Gaussian, result in reduced $\chi^2$ values. The $\chi^2$ reduction obtained with the sextupole fitting model is similar to reductions obtained with other models which include non-Gaussian modifications of the same-side 2D peak function. Such {\em local} (on azimuth) modifications only affect correlation structure at small relative azimuth and are distinct from {\em global} model elements such as higher harmonics which act over all azimuth. Although $\chi^2$ values are reduced when non-Gaussian structure is included in the fitting model, any changes in the residuals are not systematically significant. A single 2D Gaussian hypothesis for the Au+Au minimum-bias $p_t$-integral correlations from STAR is not excluded by the data.

This analysis also demonstrates that inclusion of $m > 2$ harmonics in the fitting function destabilizes the optimum parameter solutions and produces large uncertainties in the fitted parameter values. We demonstrate that a specific combination of same-side model elements, the effective ridge plus the reduced 2D Gaussian, is stable against variation of the sextupole amplitude. The present analysis demonstrates that this model-element combination is equivalent, within systematic uncertainties, to the original same-side 2D peak function obtained prior to introduction of the superfluous $m > 2$ harmonics.

Lastly, we demonstrate that for the most extreme case observed so far, the ATLAS 2.76 TeV Pb+Pb 0-1\% centrality angular correlation data with its away-side double-peaked structure, accurate phenomenological description of those data does not require a sextupole. The present analysis also shows that a non-Gaussian model is required to describe accurately the same-side 2D peak structure of those data.

In our opinion the inclusion of higher-order ($m > 2$) harmonic model elements in the description of 2D angular correlation data from relativistic heavy-ion collisions at the RHIC and the LHC has been a distraction. Attention to these superfluous higher-order harmonics rather than to the detailed structure of the same-side 2D peak has mislead efforts to understand the $\eta_\Delta$-elongated same-side 2D correlation peak and its abrupt centrality dependence. The present results should motivate the study of the evolution of non-Gaussian structure in the same-side 2D peak. Such structure may result from modified fragmentation of minimum-bias jets in heavy-ion collisions.

\vspace{0.1in}

The authors express sincere thanks to Drs. Jiangyong Jia and Peter Steinberg of the ATLAS Collaboration
for providing the data used in Sec.~\ref{SecVI}. 
This work was supported in part by the Office of Science of
The United States Department of Energy under grants No. DE-FG02-94ER40845 (UTA)
and DE-FG03-97ER41020 (UW).

%%%%%%%%%%%%%%%%%%%%%%%%%%%%%%%%%%%%%%%%%%%%%%%%%%%%%%%%%%%%%%%%%%%%%%%%%%%


\begin{thebibliography}{99}
%%%%%%%%%%%%%%%%%%%%%%%%%%%%%%%%%%%%%%%%%%%%%%%%%%%%%%%%%%%%%%%%%%%%%%%%%%%

\bibitem{aya}
J. Adams {\em et al.} (STAR Collaboration),
Phys. Rev. C {\bf 73}, 064907 (2006).

\bibitem{axialCI}
G. Agakishiev {\em et al.} (STAR Collaboration),
Phys. Rev. C {\bf 86}, 064902 (2012).

\bibitem{Joern}
B. Abelev {\em et al.} (STAR Collaboration),
Phys. Rev. C {\bf 80}, 064912 (2009).

\bibitem{atlas}
G. Aad {\em et al.} (ATLAS Collaboration),
Phys. Rev. C {\bf 86}, 014907 (2012).

\bibitem{cms}
S. Chatrchyan {\em et al.} (CMS Collaboration),
Eur. Phys. J. C {\bf 72}, 2012 (2012).

\bibitem{alice}
K. Aamodt {\em et al.} (ALICE Collaboration),
Phys. Lett. B {\bf 708}, 249 (2012);
Phys. Rev. Lett. {\bf 107}, 032301 (2011);
A. Timmins (ALICE Collaboration),
J. Phys. G: Nucl. Part. Phys. {\bf 38}, 124093 (2011).

\bibitem{Tomjetfrag}
T. A. Trainor and D. T. Kettler,
Phys. Rev. C {\bf 83}, 034903 (2011).

\bibitem{Tommodfrag}
T. A. Trainor, Phys. Rev. C {\bf 80}, 044901 (2009).

\bibitem{PHOBOSridge}
B. Alver {\em et al.} (PHOBOS Collaboration),
Phys. Rev. Lett. {\bf 104}, 062301 (2010).

\bibitem{AlverRoland}
B. Alver and G. Roland,
Phys. Rev. C {\bf 81}, 054905 (2010).

\bibitem{PHOBOSdata}
B. Alver {\em et al.} (PHOBOS Collaboration),
Phys. Rev. C {\bf 81}, 024904 (2010).

\bibitem{STARdata}
B. Abelev {\em et al.} (STAR Collaboration),
arXiv:0806.0513v1 (2008).

\bibitem{AMPT}
Z.-W. Lin, C. M. Ko, B.-A. Li, B. Zhang and S. Pal,
Phys. Rev. C {\bf 72}, 064901 (2005).

\bibitem{Tomv3-1}
T. A. Trainor, arXiv:1109.2540v1 (2011).

\bibitem{Tomv3-2}
T. A. Trainor, D. J. Prindle and R. L. Ray,
Phys. Rev. C {\bf 86}, 064905 (2012).

\bibitem{vmHydro}
L. Adamczyk {\em et al.} (STAR Collaboration),
Phys. Rev. C {\bf 88}, 014904 (2013);
B. Schenke, S. Jeon and C. Gale,
Phys. Rev. C {\bf 85}, 024901 (2012);
R. Lacey (PHENIX Collaboration), J. Phys. G: Nucl. Part. Phys. {\bf 38}, 124048 (2011).

\bibitem{HBT}
U. A. Wiedemann and U. Heinz,
Phys. Rep. {\bf 319}, 145 (1999).

\bibitem{Pearson}
J. L. Rodgers and W. A. Nicewander, Am. Stat. {\bf 42}, 59 (1988);
B. S. Everitt and A. Skrondal, {\em The Cambridge Dictionary of Statistics},
4th ed. (Cambridge University Press, Cambridge, 2010), p. 107.

\bibitem{yt}
Transverse rapidity is defined as $y_t = \log[(p_t + m_t)/m_0]$ where $m_t = \sqrt{m_0^2 + p_t^2}$ is transverse mass. Constant $m_0$ is assumed to be the pion mass 0.14~GeV/$c^2$ in this paper.

\bibitem{Porter}
R. J. Porter and T. A. Trainor (STAR Collaboration),
J. Phys.: Conf. Series {\bf 27}, 98 (2005); PoS {\bf CFRNC2006}, 004 (2006).

\bibitem{STARHBT}
J. Adams {\em et al.} (STAR Collaboration),
Phys. Rev. C {\bf 71}, 044906 (2005).

\bibitem{DavidHQ}
D. Kettler (STAR Collaboration),
J. Phys.: Conf. Ser. {\bf 270}, 012058 (2011).

\bibitem{Dihadron1}
M. M. Aggarwal {\em et al.} (STAR Collaboration),
Phys. Rev. C {\bf 82}, 024912 (2010).

\bibitem{Dihadron2}
L. Adamczyk {\em et al.} (STAR Collaboration),
arXiv:1302.6184v1 (2013).

\bibitem{Dihadron3}
A. Adare {\em et al.} (PHENIX Collaboration),
Phys. Rev. Lett. {\bf 104}, 252301 (2010).

\bibitem{etadipole}
In Sec.VII C of Ref.~\cite{axialCI} an $\eta_\Delta$-dependent modulation of the away-side dipole element was introduced to reduce the small residuals from the standard model function descriptions of the more-central Au+Au collision data. That modulation reduced the residuals over all azimuth and together with a 2D Gaussian provide an additional example of a non-Gaussian model for the same-side 2D peak.

\bibitem{FriarNegele}
J. L. Friar and J. W. Negele, in {\em Advances in Nuclear Physics}, ed. M. Baranger and E. Vogt (Plenum, New York, 1975), Vol. 8, p. 219.

\bibitem{STARTPC}
K. H. Ackermann {\em et al.} (STAR Collaboration),
Nucl. Inst. Meth. A {\bf 499}, 624 (2003); see other STAR papers in volume A {\bf 499}.

\bibitem{BW}
N. Borghini and U. A. Wiedemann, hep-ph/0506218 (2005).

\bibitem{ATLASspec}
G. Aad {\em et al.} (ATLAS Collaboration),
Phys. Lett. B {\bf 710}, 363 (2012).

\end{thebibliography}
\end{document}